\documentclass[conference]{IEEEtran-3}
\ifCLASSINFOpdf
\else
\fi

\usepackage{verbatim}
\usepackage{graphicx}
\usepackage{textcomp}
\usepackage{color}
\usepackage{breakurl}
\usepackage{multirow}
\usepackage{diagbox}
\usepackage{makecell}
\usepackage{booktabs}
\usepackage[normalem]{ulem}

\usepackage{enumitem}
\usepackage{subfigure}
\usepackage{xspace}
\usepackage{hyperref}
\usepackage{cleveref}
\usepackage{xcolor}
\usepackage{threeparttable}
\usepackage{tablefootnote}
\usepackage{colortbl} 
\usepackage{xcolor,pifont}
\usepackage{bbding}

\newcommand{\etal}{{\textit{et al. }}}
\newcommand{\major}[1]{\textcolor{black}{\textbf #1}}

\newcommand*\colourcheck[1]{%
  \expandafter\newcommand\csname #1check\endcsname{\textcolor{#1}{\ding{52}}}%
}
\newcommand*\colourcross[1]{%
  \expandafter\newcommand\csname #1cross\endcsname{\textcolor{#1}{\ding{56}}}%
}
\colourcheck{black}
\colourcheck{gray}
\colourcross{black}
\colourcross{gray}

\begin{document}
%
\title{ReDAN: An Empirical Study on Remote DoS Attacks against NAT Networks}

\author{
    \IEEEauthorblockN{Xuewei Feng\IEEEauthorrefmark{1}, Yuxiang Yang\IEEEauthorrefmark{1}, Qi Li\IEEEauthorrefmark{1}\IEEEauthorrefmark{2}, Xingxiang Zhan\IEEEauthorrefmark{2}, Kun Sun\IEEEauthorrefmark{3}, Ziqiang Wang\IEEEauthorrefmark{4}, \\Ao Wang\IEEEauthorrefmark{4}, Ganqiu Du\IEEEauthorrefmark{5}, Ke Xu\IEEEauthorrefmark{1}\IEEEauthorrefmark{2}\Envelope}

    \IEEEauthorblockA{\IEEEauthorrefmark{1}Tsinghua University, \IEEEauthorrefmark{2}Zhongguancun Laboratory, \IEEEauthorrefmark{3}CSIS, George Mason University}
    
    \IEEEauthorblockA{\IEEEauthorrefmark{4}Southeast University, \IEEEauthorrefmark{5}China Software Testing Center}

    
    
    
    
    
    fengxw06@126.com, yangyx22@mails.tsinghua.edu.cn, qli01@tsinghua.edu.cn, zhansingsong@gmail.com, \\ksun3@gmu.edu, \{ziqiangwang, wangao\}@seu.edu.cn, duganqiu@cstc.org.cn, xuke@tsinghua.edu.cn
    
}

\IEEEoverridecommandlockouts
\makeatletter\def\@IEEEpubidpullup{6.5\baselineskip}\makeatother
\IEEEpubid{\parbox{\columnwidth}{
		Network and Distributed System Security (NDSS) Symposium 2025\\
		24-28 February 2025, San Diego, CA, USA\\
		ISBN 979-8-9894372-8-3\\
		https://dx.doi.org/10.14722/ndss.2025.230972\\
		www.ndss-symposium.org
}
\hspace{\columnsep}\makebox[\columnwidth]{}}

\maketitle

\begin{abstract}
%
%
%
%
%
In this paper, we conduct an empirical study on remote DoS attacks targeting NAT networks (\major{ReDAN, short for Remote DoS Attacks targeting NAT}).
We show that Internet attackers operating outside local NAT networks possess the capability to remotely identify a NAT device and subsequently terminate TCP connections initiated from the identified NAT device to external servers.
Our attack involves two steps.
First, we identify NAT devices on the Internet by exploiting inadequacies in the Path MTU Discovery (PMTUD) mechanism within NAT specifications. This deficiency creates a fundamental side channel that allows Internet attackers to distinguish if a public IPv4 address serves a NAT device or a separate IP host, aiding in the identification of target NAT devices.
Second, we launch a remote DoS attack to terminate TCP connections on the identified NAT devices.
While recent NAT implementations may include protective measures, such as packet legitimacy validation to prevent malicious manipulations on NAT mappings, we discover that these safeguards are not widely adopted in real world.
%
Consequently, attackers can send crafted packets to deceive NAT devices into erroneously removing innocent TCP connection mappings, thereby disrupting the NATed clients to access remote TCP servers.
Our experimental results reveal widespread security vulnerabilities in existing NAT devices. After testing 8 types of router firmware and 30 commercial NAT devices from 14 vendors, we identify vulnerabilities in 6 firmware types and 29 NAT devices that allow off-path removal of TCP connection mappings.
%
%
Moreover, our measurements reveal a stark reality: 166 out of 180 (over 92\%) tested real-world NAT networks, comprising 90 4G LTE/5G networks, 60 public Wi-Fi networks, and 30 cloud VPS networks, are susceptible to exploitation.
We responsibly disclosed the vulnerabilities to affected vendors and received a significant number of acknowledgments. Finally, we propose our countermeasures against the identified DoS attack.


%
%
%
%

\end{abstract}


%
\IEEEpeerreviewmaketitle

\section{Introduction}
\label{sec:intro}
Network Address Translation (NAT) is a popular technique to map IP addresses between a private realm and the public realm, thus enabling hosts within a private network to transparently access hosts in the external  network~\cite{rfc2663,rfc3022}.  
Due to space exhaustion of IPv4 addresses, NAT is widely used in various network scenarios, e.g., 4G LTE/5G networks, cloud VPS networks, public Wi-Fi networks, and IoT networks,
to condense multiple local private addresses into a public one. 
%
%
According to CAIDA's investigations, more than 23\% Autonomous Systems (ASes) use NAT to conserve public IPv4 addresses and the proportion keeps increasing~\cite{livadariu2018inferring}.
Moreover, it is widely believed that NAT offers enhanced security~\cite{smith2002network,nat-geek,nat-checkpoint,rfc4787}, since NAT serves as an added security measure for private networks by concealing the actual IP addresses of internal hosts. This prevents direct exposure of the internal hosts to Internet attackers. 


%

In this paper, we undertake a comprehensive empirical study to demonstrate that real-world NAT implementations may exhibit vulnerabilities, which can be exploited by off-path attackers on the Internet to pose a substantial threat to end-to-end communication connectivity.
Particularly, by exploiting these vulnerabilities in various NAT devices (e.g., NAT gateways in public Wi-Fi networks or PDN gateways/UPF devices in 4G LTE/5G networks), we demonstrate that off-path attackers operating outside local NAT networks can launch remote DoS attacks against the NAT network (i.e., the network segment linked to the Internet through the NAT device) to cut off TCP connections initiated by the NATed clients to an external server.
This identified DoS attack can occur even when the internal NATed clients have a robust TCP/IP implementation and are free from DoS vulnerabilities. Our attack consists of two main steps, namely, i) identifying NAT devices on the Internet and ii) remotely severing TCP connections on the NAT devices.

We reveal that NAT specifications~\cite{rfc3022,rfc5382} inadequately address the Path MTU Discovery (PMTUD) mechanism~\cite{rfc1191,rfc1981}, thus creating a side channel exploitable by off-path attackers on the Internet. This side channel allows the attackers to distinguish whether a public IPv4 address belongs to a NAT device with multiple clients or a separate IP host, i.e., pinpointing NAT devices on the Internet.
As per PMTUD specifications, the path MTU value is maintained at the IP layer of the originator, limiting packet sizes (e.g., TCP, ICMP, UDP) sent to the destination. However, in NAT networks, ICMP packets generated by NAT devices may not align with the path MTU value maintained by Internal clients, despite originating from the same source (i.e., sharing the same public source IP address). This misalignment results in desynchronization and information leakage. 
By observing disparities in the sizes of the received TCP and ICMP packets, a remote server can discern whether the client is a private host within a NAT network or a separated IP host. Consequently, the server can pinpoint NAT devices with public IP addresses matching the source IP addresses of the received packets.

Upon identifying NAT devices on the Internet, we proceed to initiate our remote DoS attack, thereby terminating TCP connections originating from the NATed clients behind these devices.
We discover that, despite the incorporation of specific protective measures into recent NAT implementations, such as validating the legitimacy of received \texttt{RST} packets and preventing malicious removal of NAT mappings for corresponding TCP connections (e.g., in the NAT implementation of Netfilter within Linux 5.1 and beyond~\cite{netfilter}), these security measures have not been widely adopted in various real-world downstream NAT devices, including NAT gateways in public Wi-Fi networks, 4G LTE/5G networks, and cloud networks.
%
%
As a result, off-path attackers on the Internet may send crafted \texttt{RST} packets without an exact sequence number to the identified NAT device, tricking it into erroneously removing the maintained mappings for TCP connections. This, in turn, effectively disrupts the NATed clients' access to external servers and facilitates a remote DoS attack.

We conduct extensive evaluations on our attack. 
%
%
%
%
We first conduct end-to-end evaluations to assess the identified vulnerabilities.
The side channel for identifying NAT devices, stemming from inadequate PMTUD considerations in NAT specifications, affects all NAT implementations in our tests, including those within 6 types of native OSes, 8 types of router firmware, and 30 commercial NAT routers.
\major{By comparing with prior works~\cite{WebRTC,WebRTC-shark}} in 21 different network configurations, we demonstrate that our identification method excels.
%
Regarding the vulnerability of removing NAT mappings via crafted \texttt{RST} packets, we investigate NAT implementation disparities among native OSes, various router firmware, and commercial routers.
%
%
We show that some NAT implementations within native OSes (e.g., Netfilter in Linux 5.1 and beyond) may have defenses, whereas others (e.g., FreeBSD), particularly the majority of the router firmware types we tested (6 out of 8) and commercial NAT routers (29 out of 30) remain vulnerable.
Moreover, we demonstrate that, using a bandwidth of 5.72MBps, an off-path attacker can block all internal clients behind the vulnerable NAT device from connecting to a remote SSH server or downloading files from an FTP server.

In addition to end-to-end evaluations, we conduct real-world assessments to demonstrate that our attack could cause significant damage on the Internet.
Over an 11-month period, we identify more than 7,600 public IPv4 addresses used by NAT devices on the Internet by leveraging the identified side channel in PMTUD. These identified NAT devices are distributed across 1,289 Autonomous Systems (AS) in 124 countries around the world.
Besides, we conduct evaluations on 180 actual NAT networks, including various NAT scenarios such as public Wi-Fi networks, 4G LTE/5G networks, and cloud networks. The experimental results show that out of the 180 NAT networks, 166 are vulnerable to our attacks, causing a vulnerable proportion of more than 92\%. 

%

Finally, we present potential countermeasures. We propose enhancing NAT specifications to fix the side channel issue. This involves requiring NAT devices to both translate path MTU update messages to internal clients and synchronize their own path MTU values based on these messages. This measure ensures the consistency of the path MTU value from the same source IP address, effectively preventing information leakage and thwarting attackers' attempts to identify NAT devices on the Internet. Furthermore, we recommend that NAT devices implement stricter legitimacy checks on received TCP packets.
Particularly, verifying the sequence number of the received \texttt{RST} packets is one possible countermeasure to foil the attacker's removal on the preserved session mappings for TCP connections, thus throttling the identified DoS attack. Our prototype based on OpenWrt 22.03 confirms the effectiveness of the countermeasure.
%

\noindent \textbf{Contributions}. Our main contributions are as follows:
\begin{itemize}
[leftmargin=*]

\item We unveil a fundamental side channel in NAT specifications that can be exploited to identify NAT devices on the Internet, causing information leakage of NAT networks.
	
\item We investigate NAT implementation disparities among native OSes, various router firmware, and commercial routers. Our study shows that NAT implementations within downstream router firmware and commercial routers are prone to manipulation of TCP connection mappings, resulting in a DoS attack.

\item Our empirical study reveals that over 92\% (166 out of 180) of real-world NAT networks, including public Wi-Fi networks, 4G LTE/5G networks, and cloud networks, are vulnerable to the identified DoS attack.

\item We analyze the root cause and propose countermeasures to throttle the attack. The experiments confirm the effectiveness of our countermeasures.

\end{itemize}

\noindent 
\textbf{Ethical Considerations.} 
In this paper, we perform two types of experiments on the Internet to validate the feasibility and significance of our attacks in the real world. Specifically, we focus on identifying NAT devices on the Internet and discovering real-world NAT networks vulnerable to our DoS attack (see \S\ref{sec:real-world} for more details about the two types of experiments). Ethical considerations are given top priority during the experimentation process.
First, during identifying NAT devices on the Internet, we deploy 7 vantage points (HTTP servers) around the world. Whenever one of our vantage points receives an HTTP request, we use our method (see \S\ref{sec:identifying}) to determine whether the request is from a separate IP host or a NATed host. First of all, we obtained the requester's approval before conducting the identification. We state the purpose and the details of our experiment clearly in the HTTP page. The experiment will only proceed after the requester agrees.
Specifically, our identification will not pose any security risks to the requester.
\major{We only analyze the requester’s packet size and record the requester’s source IP address. These data are handled with the utmost care.}
The only side effect of our experiment is that we will modify the path MTU value of the requester—this adjustment does not impact the network segment where the requester resides, as the ICMP message used to modify the path MTU value is only received by the requester. Moreover, after the experiment, we mitigate this side effect by sending another ICMP message to restore the requester's path MTU value preserved for our vantage point.

Second, while discovering vulnerable NAT networks in the real world, we only test TCP connections under our control. After the approval of accessing the target NAT network, we set our machine as the TCP client and rent a VPS of ALICLOUD to act as the remote TCP server. We check whether our TCP connection will be terminated due to the attack proposed in \S\ref{sec:DoS}. If the connection is affected, we infer that the NAT device is vulnerable and thus the NATed network would be impacted. The experiments only impact our own TCP connection and do not affect normal users. We also reported the experimental results to the network administrators.
Additionally, we responsibly disclosed the identified vulnerabilities to the affected vendors and Internet Service Providers (ISPs). Our efforts have been met with a great deal of acknowledgments (see \S\ref{subsec:countermeasure} for more details).

\noindent \textbf{Availability.}
\major{The Proof of Concept (PoC) code for identifying NAT scenarios and verifying whether a NAT device is vulnerable to our DoS attack is available at \url{https://github.com/Internet-Architecture-and-Security/Remote-DoS-Attacks-against-NAT-Networks}. This repository includes detailed instructions to reproduce our attack.
}

\section{Background}
\label{sec:background}

\major{In this section, we first briefly review the NAT technique. Then, we introduce the Path Maximum Transmission Unit Discovery (PMTUD) mechanism, which will be exploited in our attack to identify NAT devices on the Internet.} 

\subsection{Network Address Translation}\label{bg:NAT}

Network Address Translation (NAT) is a popular technique allowing hosts (or mobile devices) with private IP addresses to communicate with hosts outside their local network. 
%
Through the use of a single publicly routable IPv4 address (e.g., 6.6.6.6), a local network segment behind a NAT gateway (routing devices performing NAT) can enable multiple clients (e.g., $client_2$, $client_3$, and $client_4$) to access Internet servers, thereby conserving the limited availability of public IPv4 addresses.
In essence, NAT works by translating the source IP address of packets leaving the local network to the public IP address of the NAT device that connects the local network to the Internet~\cite{rfc2663,rfc3022,rfc5382}.
%
%
When the server on the Internet responds, the response packet will be first routed to the public IP address of the NAT device.
The NAT device then translates the destination IP address of the response packet to the private IP address of the internal clients that originally sent the request packet, and finally forwards the packet to that client.
NAT offers multiple significant benefits, e.g., IP address conservation, improved security by hiding the internal network from the outside world, flexibility in network design and administration.


The key to NAT's normal operation is that the NAT device maintains a session mapping table.
%
%
When an internal client initiates a session to a remote server, the NAT device will create a session mapping in its cache table to keep track of the session. This mapping typically includes information such as the source IP address and port of the local client, the destination IP address and port of the remote server at the external realm, and the protocol being used for the session (e.g., TCP, UDP).
As packets flow between the internal client and the remote server, the NAT device updates the session mapping in its cache table to keep track of the state of the session.
In particular, the session mapping for a TCP connection will change as the state of the TCP connection changes. At the beginning, when the internal client issues a TCP \texttt{SYN} packet to initiate a connection with the remote server, the NAT device creates a session mapping for that connection and marks the state of the mapping as \texttt{SYN\_SENT}. As different TCP packets (e.g., the resulting inbound \texttt{SYN/ACK} packet, the subsequent outbound \texttt{ACK} packet, the TCP \texttt{RST} packet,  and the TCP \texttt{FIN} packet) are received, the state of the mapping will be altered accordingly to translate and forward the packets~\cite{rfc5382}.
%
In this paper, we show that real-world NAT devices often omit the validation of the received TCP \texttt{RST} packets, enabling attackers to manipulate the device's mapping state and construct a remote DoS attack.

\begin{figure}[h]
	\vspace{-2mm}
	\begin{center}
		\includegraphics[width=0.44\textwidth]{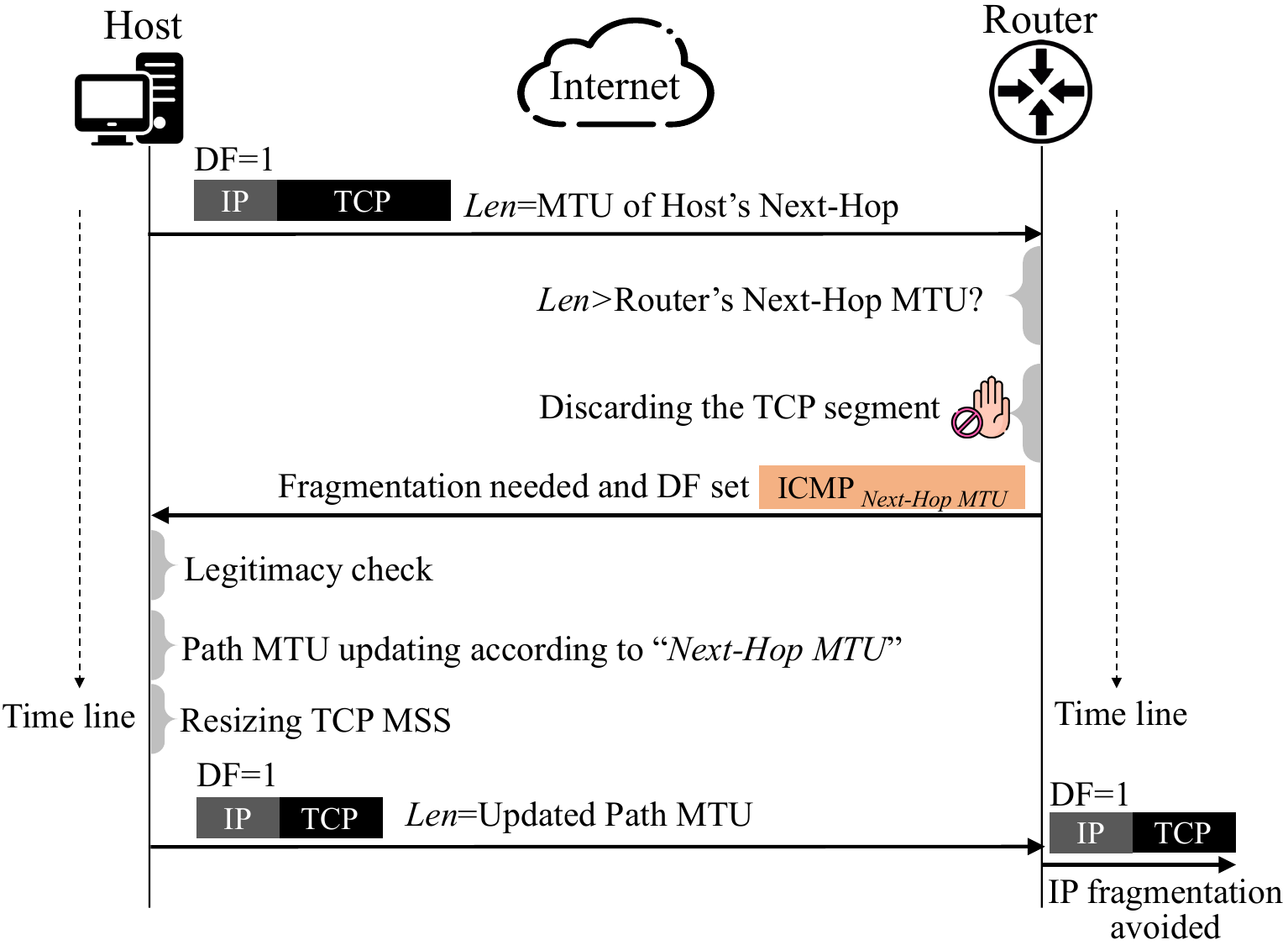}
		\vspace{-3mm}
		\caption{Workflow of PMTUD to avoid IP fragmentation.}
		\label{pic:pmtud}
	\end{center}
	\vspace{-2mm}
\end{figure}

\subsection{Path MTU Discovery}\label{bg:pmtud}

The PMTUD mechanism is a key element of the TCP/IP protocol stack~\cite{rfc1191,rfc1981,rfc8201}. PMTUD enables the efficient transmission of data over IP networks by determining the maximum packet size that can be transmitted without IP fragmentation over a particular network path, thus reducing network overhead to improve the performance.
As shown in Fig.~\ref{pic:pmtud}, PMTUD works by having the original host send packets with the Don't Fragment (DF) bit set in the IP header. 
Length of the packets is equal to the default MTU of the host's next-hop (e.g., 1500 octets in Ethernet).
If a router along the path encounters the packet that exceeds the router's next-hop MTU, it discards the packet and issues an ICMP ``Fragmentation Needed and DF Set'' message (i.e., ICMP error message with \texttt{Type}=3 and \texttt{Code}=4) back to the host, stating that the packet is too large to be transmitted. The ICMP message carries the MTU of the router's outgoing interface (i.e., next-hop MTU).

Upon receiving the ICMP error message, the host first checks the legitimacy of the message to determine whether the message is truly a reflection of the prior packet sent by itself. For example, if the message carries a TCP packet, the host will check whether the sequence number of the TCP packet is within its own sending window~\cite{feng2022ndss}.
After that, the host updates its path MTU value preserved for the destination according to the received ICMP error message. Meanwhile, the host reduces the size of the packets for the destination based on the updated path MTU value, e.g., reducing the Maximum Segment Size (MSS) of TCP packets to fit the new path MTU value and thus avoiding IP fragmentation.


According to PMTUD specifications, the path MTU value is maintained at the IP layer of the originator and thus governs all packets sent to the destination. However, in NAT networks, we discover that PMTUD's semantic consistency breaks, leading to desynchronization between internal NATed clients and the NAT device regarding path MTU values. TCP packets from internal clients adhere to the path MTU constraint, but ICMP packets from the NAT device may not, despite originating from the same source IP address. This divergence can be exploited by attackers on the Internet, potentially causing NAT network information leakage.

\section{Attack Overview}
\label{sec:overview}

\major{In this section, we present an overview of our DoS attack. First, we define the threat model of our attack. Then, we describe the steps to construct it.
}

\subsection{Threat Model}\label{sec:threat_model}

\begin{figure}[h]
	\vspace{-2mm}
	\begin{center}
		\includegraphics[width=0.44\textwidth]{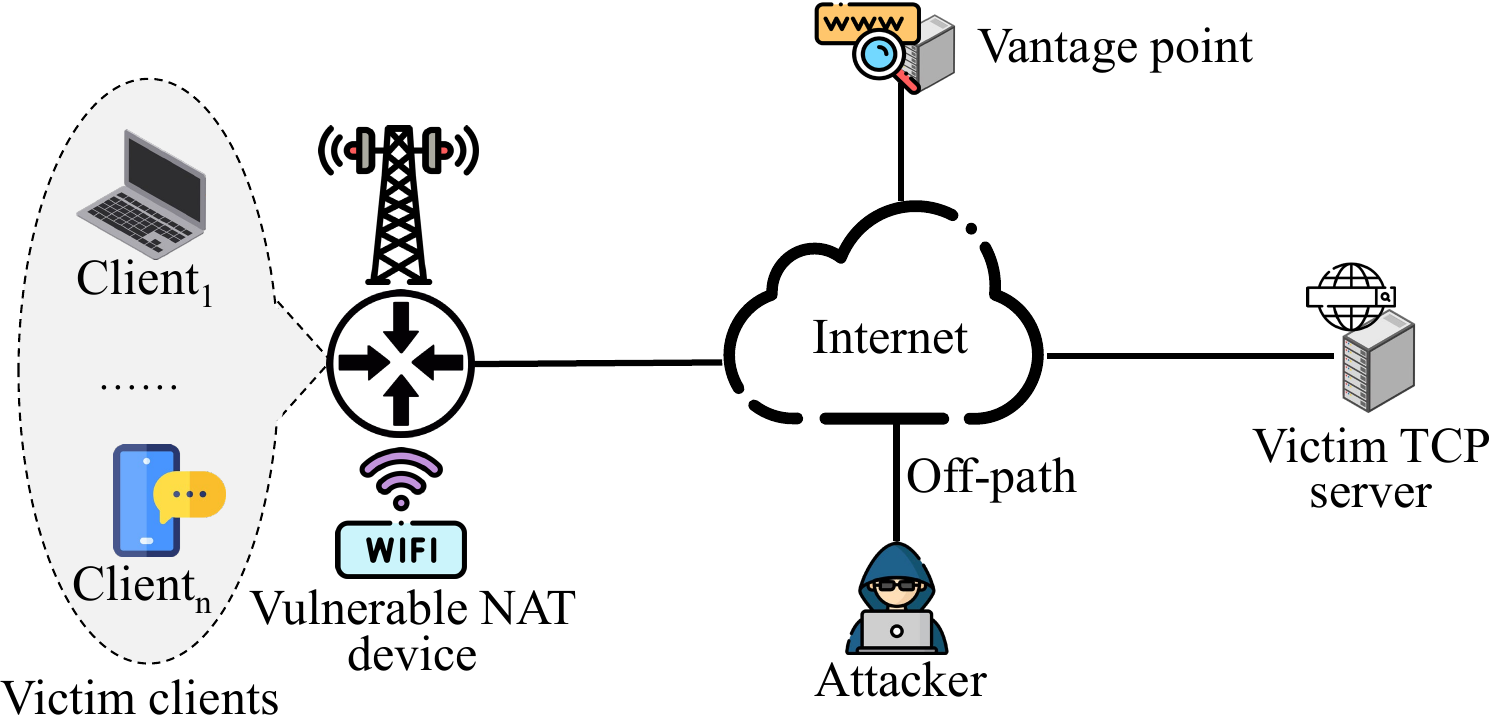}
		\caption{Threat model of remote DoS attack against NAT networks.}
		\label{pic:threat-model}
	\end{center}
	\vspace{-2mm}
\end{figure}

Fig.~\ref{pic:threat-model} shows the threat model of our off-path DoS attack. The threat model consists of 5 types of hosts.

\noindent \textit{1. A victim TCP server} is on the Internet providing various online services (e.g., popular Web services, online banking, and instant messaging) for Internet users.

\noindent
\major{\textit{2. A vulnerable NAT device} links a private network segment to the Internet. It translates the source IP address of packets leaving the private network to a publicly routable IP address, enabling access to servers in the external realm. In the context of our attack, the NAT-enabled device may be a gateway in public Wi-Fi networks, a PDN gateway or UPF in 4G LTE/5G networks, or a CPE gateway in IoT networks.}

\noindent \textit{3. Several victim clients} are located within the NATed network segment behind the vulnerable NAT device, and they access online services from the victim server.

\noindent \textit{4. An attacker} resides in the external realm, i.e., on the Internet. The attacker is capable of sending IP packets with spoofed source IP address. This capability assumption is practical, since prior studies show that about a quarter of ASes on the Internet do not filter packets with spoofed source addresses leaving their networks~\cite{luckie2019network,feng2022off-redirect}.
%
\major{The attacker aims to indiscriminately cut off TCP connections from the victim clients to the specified victim server, thereby performing a DoS attack. This attack does not target one specific client or a few select ones; instead, it affects all clients attached to the vulnerable NAT device.}

\noindent \textit{5. A vantage point} is an HTTP server deployed by the attacker on the Internet. Upon deception (e.g., via a URL-based advertisement), the victim client connects to the vantage point. The attacker can then identify whether the client is in a NAT network, potentially pinpointing a target NAT device\footnote{Our identification bypasses client-side configurations, as explained in \S\ref{subsec:NAT-identify}, without requiring JavaScript installation to access local information, which may be restricted by browser sandboxing mechanisms~\cite{WebRTC}.}.
\major{It is worth noting that in practice, the vantage point can overlap with the attacker, i.e., using the same host.}

\begin{figure*}[h]
	\begin{center}
		\subfigure[The client is a separate host without enabling NAT.]{ 
			\label{pic:identifying_no}  
			\includegraphics[width=0.46\textwidth]{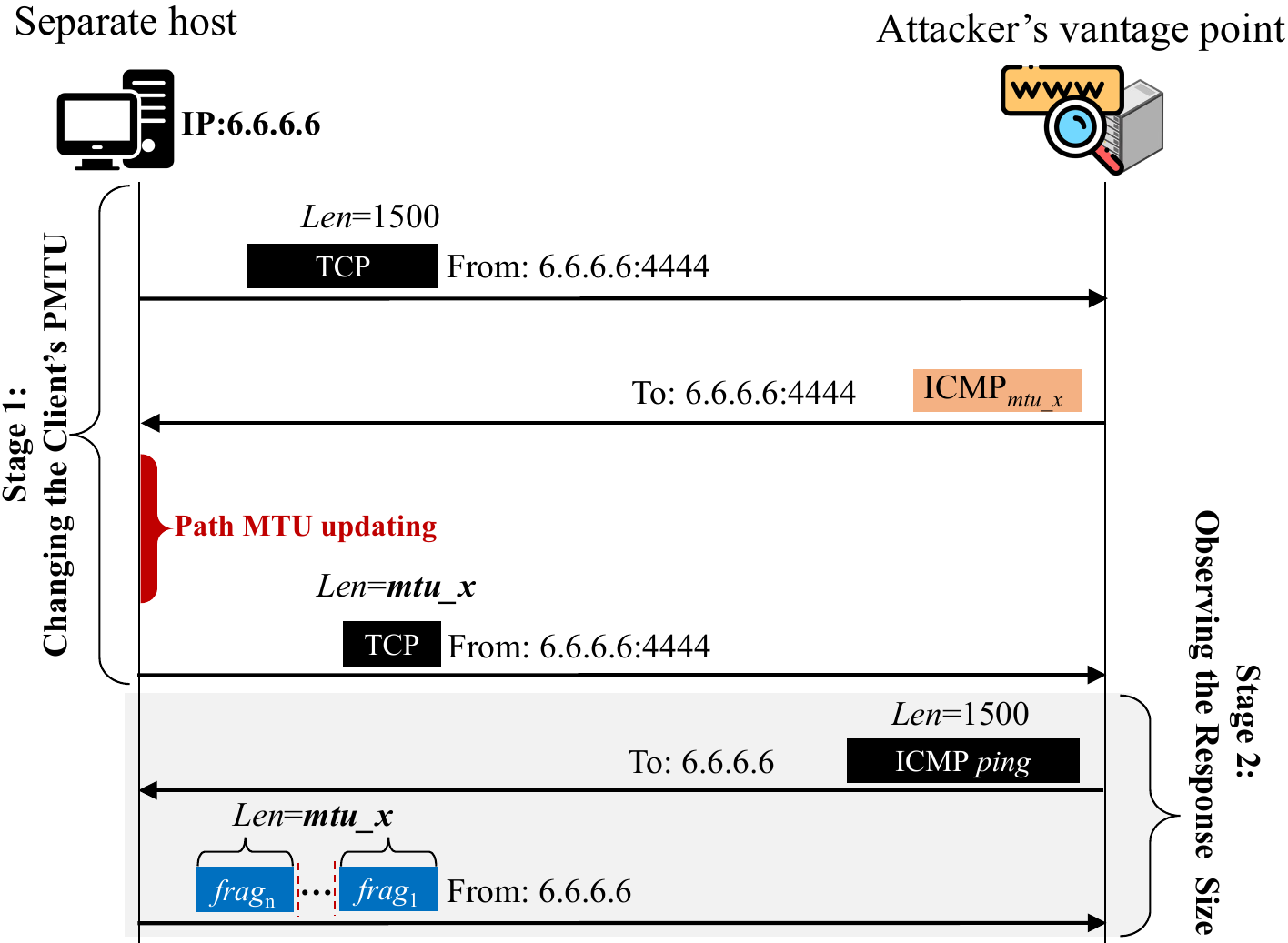} 
		} 
		\subfigure[A NAT device with public IP address of 6.6.6.6 is identified.]{ 
			\label{pic:identifing_yes}
			\includegraphics[width=0.45\textwidth]{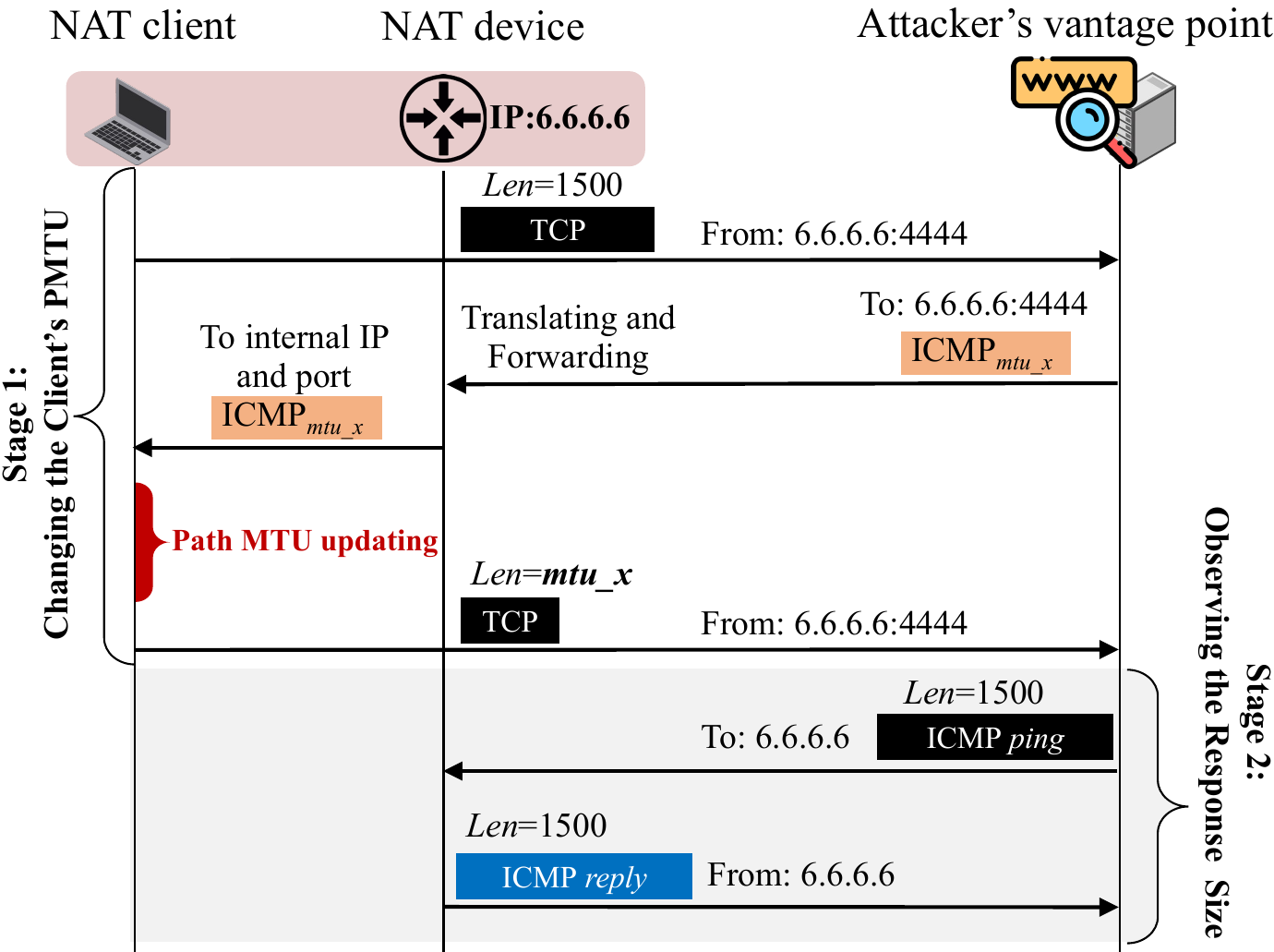}
		}
		\vspace{-1mm}
		\caption{Identifying NAT devices by leveraging a side channel presenting in the mechanism of path MTU discovery.} 
		\label{pic:identifying_NATed_Networks} 
	\end{center}
	\vspace{-6mm}
\end{figure*}

\subsection{Attack Steps}\label{sec:goal}

Our DoS attack consists of 2 steps:

\noindent \textit{Step 1. Identifying NAT Devices.}
Clients behind NAT devices, tricked by URL-based ads on social platforms or forums, may connect to our vantage point. Leveraging the side channel within the PMTUD mechanism of NAT specifications, our vantage point identifies the client within NAT networks, not as a separate IP host. This leads to the identification of a target NAT device for subsequent attacks, characterized by the client's public source IP address.

\noindent \textit{Step 2. Conducting DoS Attacks.}
By crafting \texttt{RST} packets without exact sequence numbers and issuing them to the identified NAT device, the attacker deceives the device into mistakenly removing the maintained mappings for TCP connections between the victim clients and the victim server. Subsequently, the attacker injects manipulated TCP packets into the targeted connection, creating an inconsistency between the victim clients and the victim server, effectively executing a DoS attack to terminate the connections.
\major{As TCP is a fundamental protocol of the Internet, higher-layer applications built upon it will be affected by our DoS attack, such as SSH, Web, and FTP.}

It is worth noting that in the context of the off-path threat model, the victim TCP server and destination port are typically publicly known~\cite{cao2016off,cao2018off,ccsfeng,feng2021off} (e.g., popular HTTP servers on port 80). By crafting TCP \texttt{RST} packets, the attacker can remove the identified NAT device’s session mappings (or disturb the establishment of session mappings) to such a server. Consequently, the attacker can prevent the NATed clients from accessing the specified victim server.
Besides, the attacker is not required to exactly detect a target TCP connection on the NAT device. In other words, the attacker does not rely on side channels~\cite{ccsfeng,cao2016off} to accurately infer the exact source port number (i.e., the random ephemeral port) of a target TCP connection (see \S\ref{sec:DoS} for the details of our attack). 
In practice, the source port number of a TCP connection is always within a narrow range, e.g., from 32768 to 61000 in Linux systems and from 49152 to 65535 in Windows systems~\cite{ccsfeng}. 
As a result, with just the knowledge of the public IP address of the NAT device, the attacker can utilize our attack in parallel to simultaneously cut off multiple existing TCP connections issued from these source ports or prevent the establishment of TCP connections to the specified victim server, thus constructing the DoS attack to disrupt communications between the NAT network and the victim server (refer to \S\ref{subsec:case-costs} for detailed insights into the impacts and costs associated with our DoS attack case studies).
In the next two sections, we elaborate the two attack steps one by one.

\section{Identifying NAT Devices}
\label{sec:identifying}

According to the NAT specifications~\cite{rfc2663,rfc3022,rfc5382}, servers on the Internet are unable to distinguish whether an access request was issued from a NATed client or a separate IP host.
If this were the case, there would be an information leakage of the NAT network, whereby the server would recognize that the source IP address of the received request is a publicly routable IP address of the NAT device, instead of a separate IP host.
However, we discover a side channel in the PMTUD mechanism which is inadequately addressed in NAT specifications. This side channel enables the server (e.g., an attacker's vantage point) to stealthily identify NAT devices on the Internet.
%
%
%
%
Fig.~\ref{pic:identifying_NATed_Networks} shows our method of how to identify a NAT device on the Internet.
%
%
In a nutshell, our method consists of two stages. First, the attacker's vantage point (i.e., an HTTP server) tricks the connected client into changing its path MTU value preserved for the vantage point. Second, the vantage point sends ICMP requests (i.e., ICMP ping packets) to the client and then observes the size of the replies to identify whether the client is a NATed client or not. Next, we elaborate these two stages.

\subsection{Changing Client's Path MTU}

\begin{figure}[h]
	\vspace{-2mm}
	\begin{center}
		\includegraphics[width=0.46\textwidth]{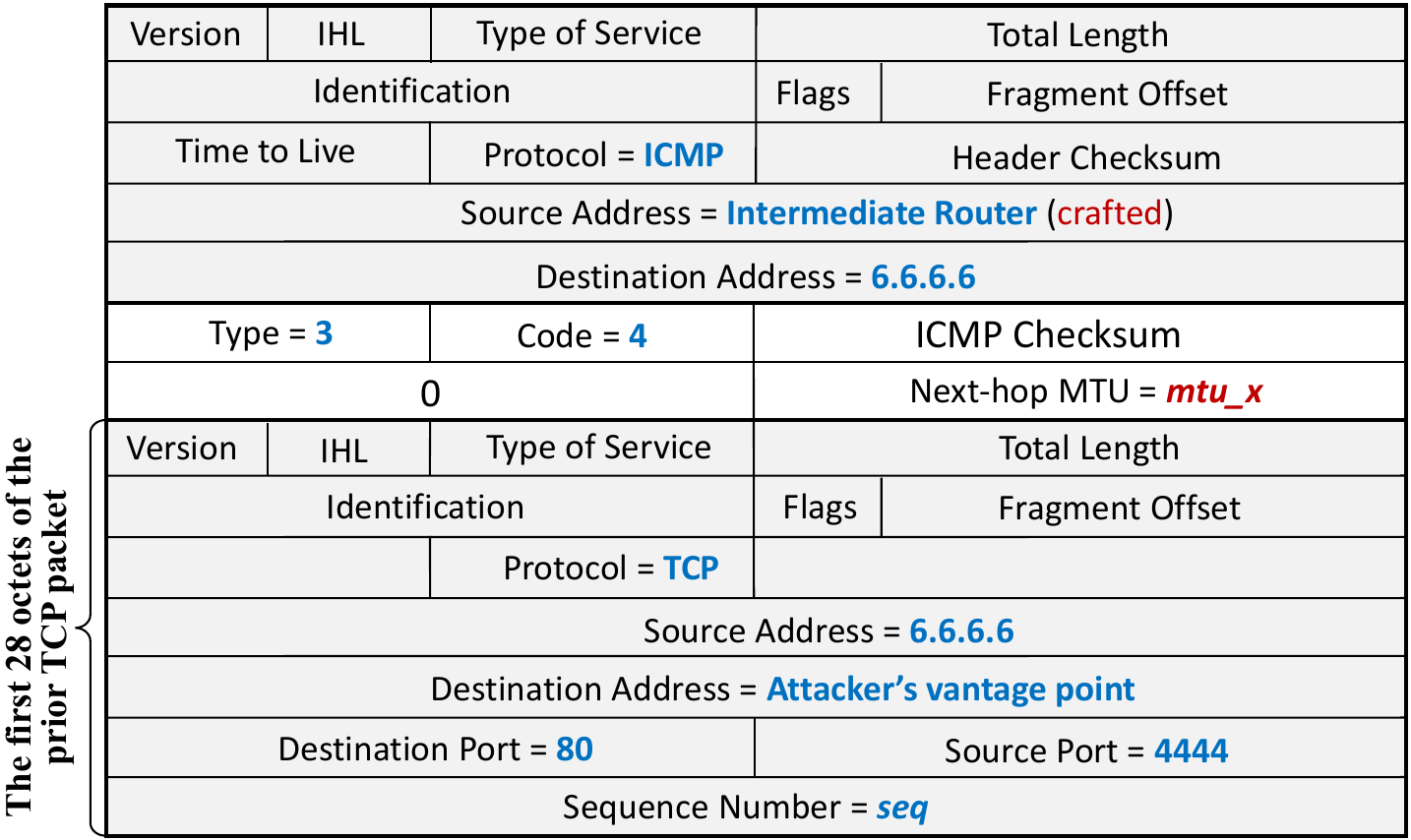}
		\vspace{-2mm}
		\caption{The crafted ICMP fragmentation needed message.}
		\label{pic:icmp}
	\end{center}
	\vspace{-4mm}
\end{figure}

%
At the beginning, the vantage point acquires the following information from the received TCP packet issued from the client: the client's IP address (e.g., 6.6.6.6), the client's source port number (e.g., 4444), the TCP sequence number of the client (i.e., \textit{seq}), and the packet size (e.g., 1500 octets).
Then, the vantage point impersonates an intermediate router and issues a crafted ICMP ``Fragmentation Needed and DF Set'' message to the client (as shown in Fig.~\ref{pic:icmp}), indicating that the prior TCP packet size exceeds the router's next-hop MTU and was discarded by the router. 
The crafted ICMP message carries the first 28 octets of the prior TCP packet, i.e., IP header of the packet, the source port number of 4444, the destination port number of 80 (i.e., the HTTP server in our example), and the sequence number of \textit{seq}.
Besides, the vantage point specifies the Next-Hop MTU value in the crafted ICMP message to \textit{mtu\_x}, which is smaller than 1500.
Note that according to the ICMP specifications, ICMP ``Fragmentation Needed and DF Set'' messages may be returned by any router on the path from the client to the vantage point. Therefore, it is difficult to verify the legitimacy of the source of this type of ICMP error messages, which means the source IP address of the message can be arbitrarily specified by the vantage point.

If the client is a separate IP host (as shown in Fig.~\ref{pic:identifying_no}), it will first perform a legitimacy check on the received ICMP ``Fragmentation Needed and DF Set'' message. The message carries information derived from the prior TCP packet, so it will pass the client's check and the client will respond to the message according to the PMTUD mechanism as mentioned prior in \S\ref{bg:pmtud}. The client will adjust the length of its subsequent TCP packets for the vantage point to \textit{mtu\_x}.
%
%
In contrast, if the client is a NATed client (as shown in Fig.~\ref{pic:identifing_yes}), the NAT device will receive the ICMP message first and then translate the message to the internal client. After receiving the ICMP message, the client will respond to the message according to the PMTUD specifications (i.e., updating its path MTU to the attacker's vantage point), and finally adjust the length of the TCP packets to \textit{mtu\_x}.
As a result, when the client sends subsequent TCP packets to the vantage point again, the vantage point will observe a significant change in the packet length, i.e., decreasing from the previous 1500 octets to \textit{mtu\_x}. The vantage point can successfully infer that the client's path MTU value has been deceitfully updated.

\subsection{Observing Response Sizes}

After changing the client's path MTU, the vantage point initiates an ICMP ping request packet (i.e., ICMP packet with \texttt{Type}=8 and \texttt{Code}=0) to the client’s public routable IP address (i.e., 6.6.6.6). The length of the ping request packet is padded to the previously observed 1500 octets from the regular TCP packets. Distinctively, the generation of the ICMP reply packet (i.e., ICMP packet with \texttt{Type}=0 and \texttt{Code}=0) significantly differs between the two scenarios of a separate host and a NAT client.

If the remote client is a separate host (as shown in Fig.~\ref{pic:identifying_no}), the ping request packet sent by the vantage point will be received directly by the host. Affected by the previously updated path MTU value (i.e., decreased to \textit{mtu\_x}), the returned 1500-octet ICMP reply packet will be fragmented into several IP fragments by the host, each with a length of \textit{mtu\_x}.
On the contrary, if the client is a NAT client, the ping request packet sent by the vantage point will arrive at the NAT device instead of being received by the client (as shown in Fig.~\ref{pic:identifing_yes}). As a result, the length of the returned ICMP reply packet issued by the NAT device will be the default 1500 octets without being fragmented\footnote{IP fragmentation at intermediate routers will not affect our identification on NAT devices. See \S\ref{subsec:NAT-identify} for our detailed evaluations on this issue.}. This is because the previous ICMP ``Fragmentation Needed and DF Set'' message sent by the vantage point updated the path MTU value of the internal NAT client, while the path MTU value preserved in the NAT device for the vantage point was not affected. Consequently, the path MTU value from the same IP address (i.e., 6.6.6.6 in our example) to the vantage point has been desynchronized, i.e., the length of the TCP packet is \textit{mtu\_x} while the length of the ICMP reply packet is 1500 octets, forming a side channel.

In summary, by tricking the client into altering its path MTU value, the attacker can differentiate between separate IP hosts and NAT clients by analyzing the ICMP reply packet lengths, enabling subtle identification of NAT devices on the Internet. If the received ICMP reply packet is not fragmented (as shown in Fig.~\ref{pic:identifing_yes}), the client is a NATed host, and the public IP address of the observed NATed client corresponds to that of the NAT device, all achieved without requiring additional assistance.
\major{Note that while the attacker can detect the presence of NATed clients and thus conduct a DoS attack against the identified NAT device, it cannot determine the exact number of clients or distinguish between individual ones. Additionally, in cascaded NAT networks (i.e., chained NAT contexts), our method can identify the NATed client accessing the vantage point and enable DoS attacks targeting the outermost NAT device hosting the public IP address. However, it cannot differentiate between the contexts of chained NATs.}

\section{Conducting DoS Attacks}
\label{sec:DoS}
%
%

After identifying NAT devices on the Internet, we proceed to conduct our remote DoS attack against these identified NAT devices to disrupt their TCP connections to a specified victim server.
According to the NAT specifications~\cite{rfc5382,rfc3022}, NAT behavior for handling TCP \texttt{RST} packets is left unspecified.
In practice, NAT implementations in native OSes (e.g., Netfilter in Linux 5.1 and beyond) may validate the legitimacy of received TCP \texttt{RST} packets before altering the state of the mappings preserved for associated TCP connections~\cite{netfilter}.
However, this validation may not be enforced in various real-world downstream router-oriented NAT devices, such as NAT gateways in public Wi-Fi networks, 4G LTE/5G networks, and cloud networks.
By exploiting this vulnerability, an off-path attacker can craft TCP \texttt{RST} packets to remove the mappings maintained in the NAT device.
This disruption will result in a DoS attack against the NAT network.
Our DoS attack consists of three stages, i.e., removing NAT mappings, manipulating TCP states, and terminating TCP connections.
%
%
%
%
%

\subsection{Removing NAT Mappings}
\begin{figure}[h]
\vspace{-2mm}
    \begin{center}
		\includegraphics[width=0.49\textwidth]{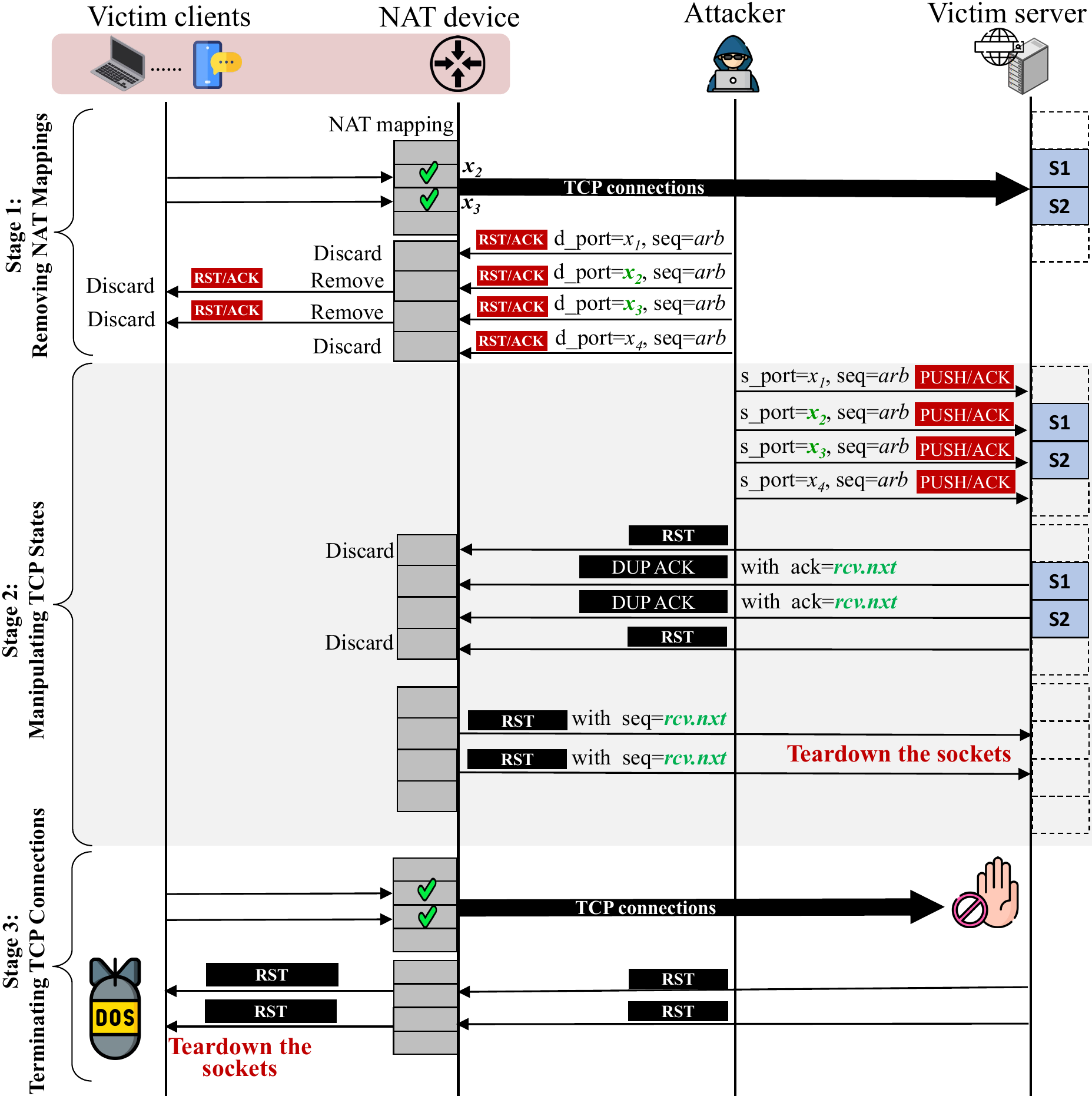}
		\vspace{-4mm}
		\caption{Design of our remote DoS attack against NAT networks.}
		\label{pic:dos}
	\end{center}
	\vspace{-3mm}
\end{figure}

As shown in Fig.~\ref{pic:dos}, at the beginning, the NAT device maintains two session mappings (using the source port $x_2$ and $x_3$) for TCP connections issued from two internal clients to the remote victim server. The server preserves two sockets, i.e., \texttt{s1} and \texttt{s2}, for the two TCP connections.
The parameters of the TCP connections (e.g., source port numbers and sequence numbers) cannot be observed by the off-path attacker on the Internet. 
The attacker impersonates the victim server to craft multiple TCP \texttt{RST} (or TCP \texttt{RST/ACK}) packets\footnote{In practice, some NAT devices, e.g., the carrier-grade NAT (CGNAT) devices within the China Unicom Beijing province network, may also examine the \texttt{ACK} flag (not the sequence number) in the received \texttt{RST} packet prior to removing TCP connection mappings. Consequently, in our empirical measurement study, we consistently set the \texttt{ACK} flag of the crafted \texttt{RST} packets to true as well (i.e., a crafted TCP \texttt{RST/ACK} packet), ensuring the success of our attack in different NAT networks.} to the identified public IP address of the NAT device. The source IP address \texttt{s\_IP} of the crafted \texttt{RST} packets is specified as the victim server’s IP address. The destination port number \texttt{d\_port} (i.e., the source port of the client-issued TCP connections) is typically located within a small range, e.g., from 32768 to 61000 in Linux systems and from 49152 to 65535 in Windows systems~\cite{ccsfeng}. The source port number (i.e., the destination port of the TCP connections) is usually known, e.g., 80 in HTTP. The sequence number \texttt{seq} of the crafted \texttt{RST} packets is arbitrary (i.e., specified by the attacker arbitrarily). The acknowledgment number is not required.
The crafted TCP \texttt{RST} packets with incorrect \texttt{d\_port} will be discarded by the NAT device directly. By contrast, the packets with correct \texttt{d\_port} (i.e., $x_2$ and $x_3$) will be translated into the internal clients by the NAT device.
Because the sequence number in the packets are incorrect, the internal clients with a robust TCP/IP protocol suite implementation will eventually discard them.
However, the NAT device may be tricked into removing the session mappings for the two TCP connections without verifying the legitimacy of the \texttt{RST} packets, primarily for performance considerations.
%

%
%

\subsection{Manipulating TCP States}
After removing the session mappings, the attacker then impersonates the NAT device (i.e., using the identified public IP address of the device) to craft multiple TCP \texttt{PUSH/ACK} packets (i.e., with the \texttt{PUSH} and \texttt{ACK} flags in TCP header enabled) to the server.
The source port number \texttt{s\_port} of the crafted \texttt{PUSH/ACK} packets is specified within the range that TCP usually selects, and the destination port number is specified as the known number in the server (e.g., 80 in HTTP). The sequence number and the acknowledgment number are arbitrary.
The crafted \texttt{PUSH/ACK} packets will trigger the server to respond differently. As shown in Fig.~\ref{pic:dos}, for \texttt{PUSH/ACK} packets with source port equal to the previously established TCP connections' source port (i.e., the source port of $x_2$ and $x_3$ in our example), the server will issue a duplicate acknowledgment packet to the public IP address of the NAT device according to the fast retransmit mechanism defined in RFC 5681~\cite{rfc5681}. 
Particularly, the duplicate acknowledgment packet carries the client's exact sequence number of \textit{rcv.nxt}, i.e., the lower boundary of the server’s receive window\footnote{We observe that servers running Linux, Windows, FreeBSD, and macOS consistently adhere to the fast retransmit mechanism defined in RFC 5681. They generate duplicate acknowledgment packets upon receiving the \texttt{PUSH/ACK} packets. In contrast, OpenBSD systems deviate from this behavior, as they do not produce the expected duplicate ACK packets. Consequently, servers equipped with OpenBSD will not be affected.}. 
By contrast, for \texttt{PUSH/ACK} packets carrying a wrong source port, the server will reflect a TCP \texttt{RST} packet to the NAT device, since no TCP sockets with those source ports are preserved on the server for the public IP address of the NAT device. 
The NAT device will discard the received \texttt{RST} packets silently.

Once the duplicate acknowledgment packets reach the NAT device, the NAT device may generate an \texttt{RST} packet to the server for each received duplicate acknowledgment packet. This occurs because the mappings for the TCP connections were previously removed due to the attacker's crafted TCP \texttt{RST} packets, leading the NAT device to believe that the corresponding TCP sockets have been terminated.
According to the TCP specifications~\cite{rfc793,rfc9293}, a TCP \texttt{RST} packet will be sent whenever a non-RST packet arrives at a closed socket.
%
Moreover, the sequence number (i.e., \texttt{seq}) of the \texttt{RST} packets issued from the NAT device will be specified as the clients' exact sequence number (i.e., \textit{rcv.nxt} of the server), copied from the previously received duplicate acknowledgment packets. 
In accordance with TCP specifications, the \texttt{RST} packets that carry the exact sequence number of the clients will deceive the server into tearing down the corresponding TCP sockets.
In practice, the attacker can optimize the attack by interleaving stage 1 (issuing crafted \texttt{RST} packets) and stage 2 (issuing crafted \texttt{PUSH/ACK} packets), thereby mitigating potential limitations in the victim client's TCP packet transmission between these stages, which could affect the attack's success (refer to \S\ref{subsec:case-costs} for experimental details in our case studies).

\subsection{Terminating TCP Connections}

Finally, once the clients have subsequent TCP packets to send to the server through the previously established TCP connections, the NAT device will first create new mappings. Once the TCP packets arriving at the server, the server will directly discard the packets. Moreover, the server will issue \texttt{RST} packets to the clients according to TCP specifications. 
These \texttt{RST} packets carry the exact sequence number (copied from the \texttt{Acknowledgment Number} field of the prior received TCP packets issued by the clients) of the connections, and eventually force the clients to tear down the TCP connections, i.e., leading to a DoS attack.

\section{Evaluations}
\label{sec:end-to-end}

In this section, we conduct a comprehensive end-to-end evaluation of our attacks. We begin with the experimental setup, followed by the evaluation results. This includes identifying NAT devices via the PMTUD side channel, testing whether NAT mappings in various implementations can be maliciously removed by attackers, and assessing the attack's costs through case studies on DoS attacks over SSH and FTP.

%
%

%
%

\begin{figure}[h]
	\begin{center}
		\includegraphics[width=0.42\textwidth]{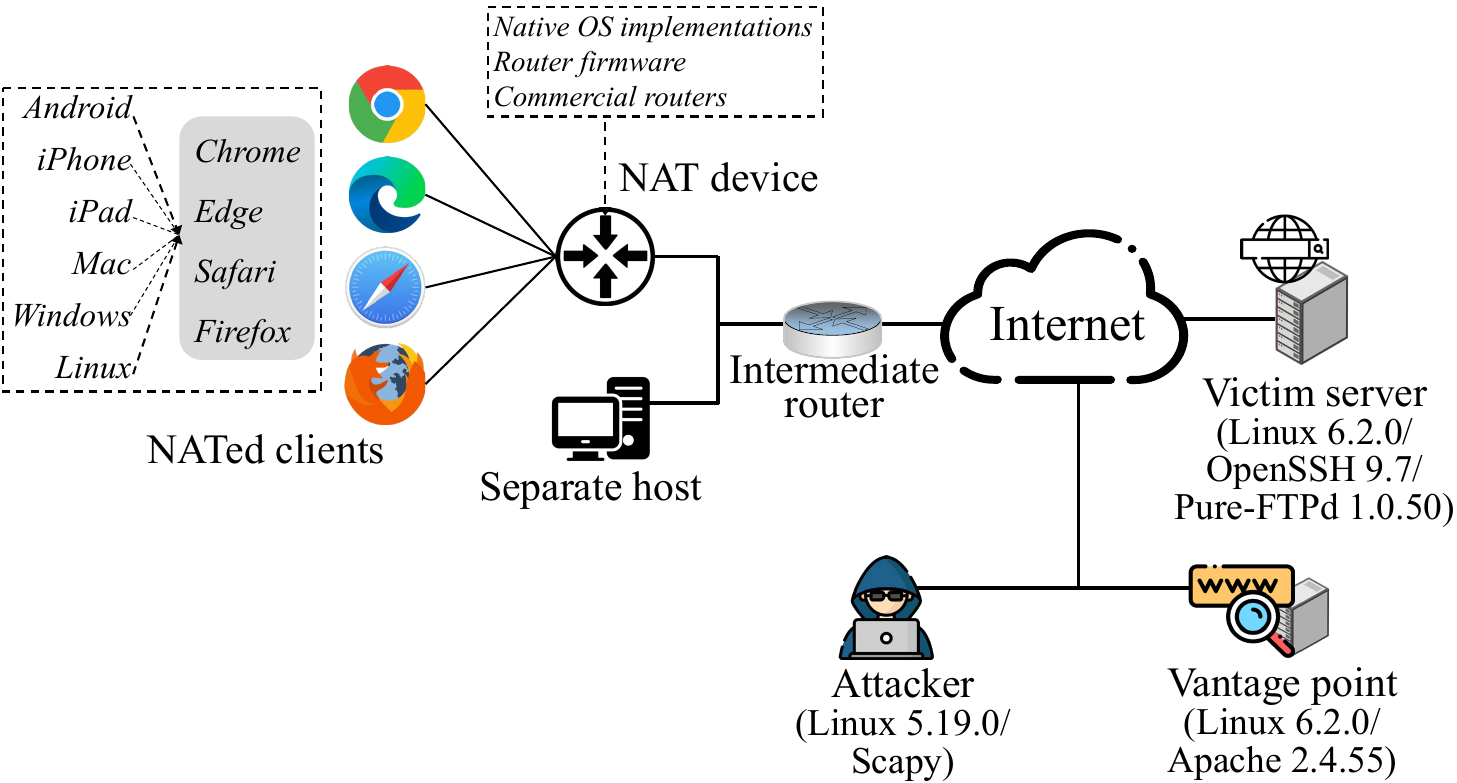}
		\vspace{-2mm}
		\caption{Experimental setup for end-to-end evaluations.}
		\label{pic:setup}
	\end{center}
	\vspace{-4mm}
\end{figure}

\subsection{Experimental Setup}\label{subsec:end-to-end-setup}

%
Fig.~\ref{pic:setup} shows the experimental setup for conducting end-to-end evaluations of our attacks.
The NAT device setup includes forwarding-enabled hosts with 6 native OSes and 8 types of router firmware implementing NAT, as well as 30 commercial routers with NAT enabled (see Table~\ref{NAT_removing} for NAT setup details). Connected NATed clients have varied configurations, using different OS and browsers. This diversity facilitates the comparison of our NAT device identification method with malicious JavaScript-based methods~\cite{WebRTC}.
The separate host shares an identical configuration with the NATed clients, the sole difference being its possession of a public IP address, allowing direct access to the vantage point. This setup is designed to evaluate the effectiveness of our NAT device identification method, specifically in its ability to distinguish between a NATed client and a separate host at the vantage point.
An intermediate router equipped with OpenBSD 7.4 is utilized to check whether IP fragmentation at intermediate routers affects our identification method.
The victim server offers SSH connectivity and FTP file download services to the NATed clients.
Our evaluation involves three tests: First, the attacker identifies NAT devices via the identified PMTUD side channel. Second, the attacker examines if NAT mappings in the devices can be manipulated for our DoS attack. Third, we carry out case studies on DoS attacks targeting SSH connections and FTP downloads from the NATed clients to the victim server, assessing the impacts and cost implications of our attack.

\subsection{NAT Devices Identification} \label{subsec:NAT-identify}

Initially, the NATed clients and the separate host separately click on a URL to access the vantage point.
Following the procedure in Fig.~\ref{pic:identifying_NATed_Networks}, the vantage point can effectively distinguish requests from the separate host and the NATed clients. In other words, all NAT setups listed in Table~\ref{NAT_removing} are affected by the side channel in the PMTUD mechanism.
The differentiation is based on the alignment between the TCP packet size of the separate host and its ICMP ping reply packet size, both adjusted to a 600-octet path MTU value (i.e., \textit{mtu\_x}) specified by the vantage point. In contrast, the NATed clients display a different behavior, with their TCP packet size adhering to the 600-octet path MTU, while the ICMP ping reply packet size (originating from the NAT device) defaults to 1500 octets.
%

We also conduct experiments to evaluate the potential impact of IP fragmentation at intermediate routers on our identification method.
%
We modify the next-hop MTU value of the intermediate router to different sizes (i.e., 1492 octets used by IEEE 802.3 and 576 octets used by X.25 networks~\cite{rfc1191}), other than the default 1500 octets, as illustrated in Table~\ref{tab:routers}. This allows us to evaluate that even when the reflected ICMP ping reply packets are routed through a path where some intermediate routers may perform IP fragmentation, our identification method remains unaffected, since the ICMP reply packet sizes observed from the separate host and the NATed client are always distinguishable.
Another potential disruption to our identification method, arising from intermediate routers, is the possibility of the fragmented ICMP reply packet from the separate host being reassembled by certain routers, thus restoring the IP fragments to the original 1500-octet size. This has the potential to mislead the vantage point, leading to an erroneous identification of the separate host as a NAT client. However, in practice, only the destination host reassembles fragments, and intermediate routers do not enforce fragment reassembly, as fragments do not always take the same routes from source to destination~\cite{frag-router1,frag-router2}. Therefore, our identification method effectively mitigates the risk of such misidentification.

\begin{table}[h]
\setlength\tabcolsep{2.5pt}
\centering
\caption{Affects from intermediate IP fragmentation.}
\vspace{-2mm}
\label{tab:routers}
\scalebox{0.89}{
\begin{tabular}{@{}clllc@{}}
\toprule
\textit{\textbf{mtu\_x}} &
  \multicolumn{1}{c}{\textbf{\begin{tabular}[c]{@{}c@{}}Next-hop MTU\\ of the router\end{tabular}}} &
  \multicolumn{1}{c}{\textbf{\begin{tabular}[c]{@{}c@{}}ICMP reply size\\ observed\\ from separate host\end{tabular}}} &
  \multicolumn{1}{c}{\textbf{\begin{tabular}[c]{@{}c@{}}ICMP reply size\\ observed\\ from NATed client\end{tabular}}} &
  \textbf{\begin{tabular}[c]{@{}c@{}}Disting-\\ uishable\end{tabular}} \\ \hline
\multirow{3}{*}{600} & 1500 (by default)       & 2*600+300      & 1500      & \blackcheck \\
                     & 1492 (greater than 600) & 2*600+300      & 1492+8    & \blackcheck \\
                     & 576 (lower than 600)    & 2*(576+24)+300 & 2*576+348 & \blackcheck \\ \hline
\end{tabular}
}
\end{table}

Moreover, we compare our NAT identification method with the JavaScript-based method~\cite{WebRTC,WebRTC-shark}, which requires NATed clients to install a malicious JavaScript to call APIs like WebRTC, enabling access to the local IP address and comparing it with the public IP address. 
This method has significant limitations due to security policies in modern browsers, which may restrict the exposure of the local IP address to JavaScript~\cite{WebRTC}.
In contrast, our method is configuration-independent, leveraging a fundamental side channel vulnerability, thereby impacting all client configurations in Fig.~\ref{pic:setup}. Table~\ref{NAT_clients} summarizes our experimental results, showing that our method consistently identifies the NAT scenario and the corresponding public IP address across 21 client configurations\footnote{In our tests, we use browsers on various OSes, all of which are recent stable versions, e.g., Chrome 117.0.5938.60, Edge 117.0.2045.33, Safari 16.6, and Firefox 117.3.}, while the JavaScript-based method succeeds in only two configurations.

\begin{table}[h]
\centering
\caption{NAT identification with different configurations.}
\vspace{-2mm}
\label{NAT_clients}
\begin{threeparttable}
\begin{tabular}{lcccc} 
\bottomrule
 & Chrome & Edge & Safari & Firefox \\\hline
Android 12 &\blackcheck|\graycheck  & \blackcheck|\graycheck & N/A & \blackcross|\graycheck \\\hline
iOS 16.3 & \blackcross|\graycheck & \blackcross|\graycheck & \blackcross|\graycheck & \blackcross|\graycheck \\\hline
iPadOS 16.61 & \blackcross|\graycheck & \blackcross|\graycheck & \blackcross|\graycheck & \blackcross|\graycheck \\\hline
MacOS 13.0 & \blackcross|\graycheck & \blackcross|\graycheck & \blackcross|\graycheck & \blackcross|\graycheck \\\hline
Windows 10 & \blackcross|\graycheck & \blackcross|\graycheck & N/A & \blackcross|\graycheck \\\hline
Linux 6.2.0 & \blackcross|\graycheck & \blackcross|\graycheck & N/A & \blackcross|\graycheck \\
\toprule
\end{tabular}
\begin{tablenotes}
       \footnotesize
       \item[] \blackcheck means the JavaScript-based method works.
       \item[] \blackcross means the JavaScript-based method fails.
       \item[] \graycheck means our side channel-based method works.
    \end{tablenotes}
\end{threeparttable}
\end{table}

\begin{table}[h]
\centering
\renewcommand\arraystretch{0.95}
\setlength\tabcolsep{1.5pt}
\caption{TCP session mappings removal via crafted \texttt{RST}.}
\vspace{-2mm}
\label{NAT_removing}
\begin{threeparttable}
\scalebox{0.96}{
\begin{tabular}{@{}c|llcc@{}}
\toprule
\multicolumn{1}{c|}{\textbf{NAT Setup}} &
  \multicolumn{1}{l}{\textbf{\begin{tabular}[c]{@{}c@{}}OS/Firmware\\ /Router\end{tabular}}} &
  \multicolumn{1}{l}{\textbf{\begin{tabular}[c]{@{}c@{}}Version\\ /Vendor\end{tabular}}} &
  \multicolumn{1}{c}{\textbf{\begin{tabular}[c]{@{}c@{}}Release\\ Date$^*$\end{tabular}}} &
  \textbf{Vulnerable} \\ \midrule
\multirow{6}{*}{Native OS}          & FreeBSD            & 13.2 and earlier & 04/2023 & \blackcheck \\
                                    & Linux       & 5.0 and earlier    & 05/2019 & \blackcheck \\
                                    & Linux       & 5.1 and beyond    & 05/2019 & \blackcross \\
                                    & OpenBSD            & 5.0 and beyond   & 11/2011 & \blackcross \\
                                    & macOS       &  13.2.1   & 02/2023 & \blackcross \\
                                    & Windows            & \multicolumn{1}{l}{\begin{tabular}[c]{@{}c@{}}10\\11\end{tabular}} &  \multicolumn{1}{c}{\begin{tabular}[c]{@{}c@{}}07/2015\\10/2021\end{tabular}}& \blackcross \\\midrule
\multirow{8}{*}{\begin{tabular}[c]{@{}c@{}}Router \\ Firmware \end{tabular}}    & OpenWrt            & 22.03 and earlier  & 05/2023 & \blackcheck \\
                                    & AsusWrt            & 3.0.0.4.386 and earlier  & 10/2022 & \blackcheck \\
                                    & pfSense            & 2.7.0 and earlier  & 06/2023 & \blackcheck \\
                                    & OPNsense           & 23.7 and earlier  & 07/2023 & \blackcheck \\
                                    & iKuai              & 3.7.6 and earlier   & 09/2023 & \blackcheck \\
                                    & VxWorks            & 5.5.1   & 09/2002 & \blackcheck \\
                                    & VyOS               & 1.4 and beyond     & 11/2020 & \blackcross \\
                                    & RouterOS           & 6.49 and beyond    & 08/2021 & \blackcross \\ \midrule
\multirow{30}{*}{\begin{tabular}[c]{@{}c@{}}Commercial \\ Router\end{tabular}} & RAX20              & Netgear & 10/2020 & \blackcheck \\
                                    & RAX50              & Netgear & 02/2020 & \blackcheck \\
                                    & E5600              & Linksys & 03/2020 & \blackcheck \\
                                    & E9450              & Linksys & 05/2022 & \blackcheck \\
                                    & RT-AX57            & ASUS    & 02/2023 & \blackcheck \\
                                    & RT-AX89X           & ASUS    & 10/2020 & \blackcheck \\
                                    & AR6140E-9G-2AC     & Huawei  & 05/2023 & \blackcheck \\
                                    & AX3 Pro            & Huawei  & 09/2020 & \blackcheck \\
                                    & WS5200             & Huawei  & --- & \blackcheck \\
                                    & TC7102             & Huawei  & 04/2020 & \blackcheck \\
                                    & TL-R473GP-AC      & TP-Link  & 04/2021 & \blackcheck \\
                                    & TL-R4239GP        & TP-Link  & 06/2022 & \blackcheck \\
                                    & TL-XDR6020        & TP-Link  & 01/2022 & \blackcheck \\
                                    & TL-AC1200      &  TP-Link  & 12/2020 & \blackcheck \\
                                    & TL-WDR7620       &  TP-Link  & --- & \blackcheck \\
                                    & Magic R100        & H3C     & 01/2020 & \blackcheck \\
                                    & Magic R365         &  H3C   & 09/2022 & \blackcheck \\
                                    & EG105G-V2      &   Ruijie      & 05/2023 & \blackcheck \\
                                    & EG210G-P        &  Ruijie       & 01/2023 & \blackcheck \\
                                    & X32 Pro           &   Ruijie      & 08/2022 & \blackcheck \\
                                    & Redmi RA81      &   Xiaomi      & 01/2022 & \blackcheck \\
                                    & CR6609         &  Xiaomi    & --- & \blackcheck \\
                                    & NBR1009GPE     & Netcore  & --- & \blackcheck \\
                                    & MG1200AC      &  Netcore  & --- & \blackcheck \\
                                    & Wimaster         &  Wimaster   & --- & \blackcheck \\
                                    & Wimaster-mini     &  Wimaster  & --- & \blackcheck \\
                                    & Google Wi-Fi     &   Google   & 10/2016 & \blackcheck \\
                                    & SK-WR6640X       & Skyworth    & --- & \blackcheck \\
                                    & RAX1800Z         &   China Mobile  & 11/2021 & \blackcheck \\
                                    & Cisco Meraki MX64  &   Cisco Meraki & 02/2015 & \blackcross \\ \bottomrule
\end{tabular}
}
\begin{tablenotes}
       \footnotesize
       \item[] \blackcheck means the NAT implementation is vulnerable.
       \item[] \blackcross means the NAT implementation is invulnerable.
       \item[] $^*$ the release date information for the commercial routers is sourced\\ from the Internet.
    \end{tablenotes}
\end{threeparttable}
\vspace{-2mm}
\end{table}

\begin{figure}[h]
    \centering
    \includegraphics[width=0.75\linewidth]{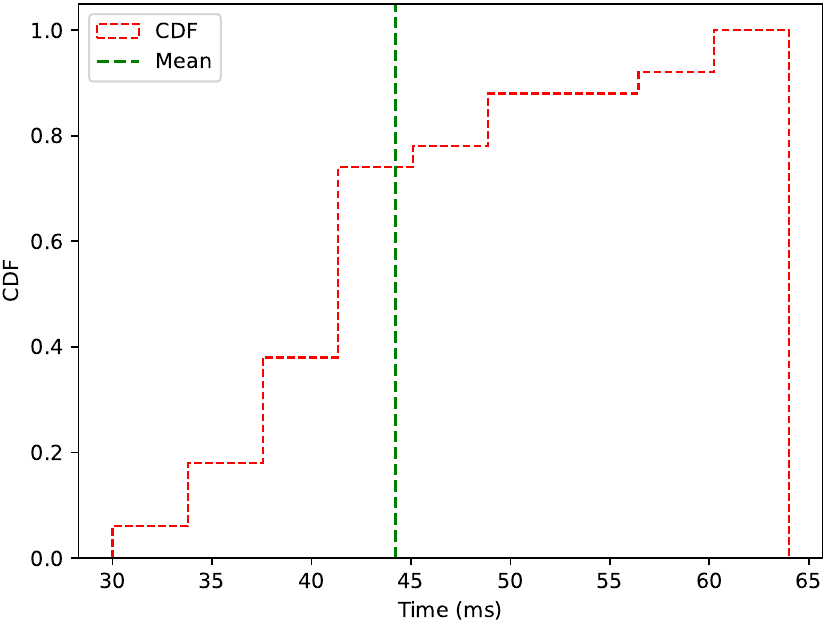}
    \vspace{-2mm}
    \caption{Time cost of identifying NAT in Wi-Fi networks through connection to common private gateway addresses.}
    \label{fig:cdf}
    \vspace{-2mm}
\end{figure}

Note that the JavaScript installed on the client may also exploit a timing side channel to identify whether NAT is enforced or not. However, this method also heavily relies on specific network configurations. The JavaScript may connect to common private gateway addresses (such as 10.0.0.1, 192.168.1.1, 192.168.0.1, etc.) through a hidden img HTML tag.
Upon successful connection, a JavaScript event is triggered, providing information that the client is a NAT client. In contrast, separate hosts will not achieve success and will raise an error after a timeout. 
We evaluate this method in two common NAT scenarios: public Wi-Fi networks and 5G cellular networks, using the same experimental setup where a client with malicious JavaScript attempts to connect to the common gateway addresses.
In 50 experiments conducted on our campus NAT-enabled Wi-Fi network, the JavaScript successfully connects to one of the 30 common private gateway addresses we listed, with an average time cost of 44 milliseconds (as shown in Fig.~\ref{fig:cdf}), effectively identifying the NAT scenario. However, in the 5 tested 5G NAT networks accessible on our campus, all experiments fail due to non-deterministic assignment of gateway addresses in 5G networks, which remains undisclosed to users.

\begin{table*}[]
\centering
\renewcommand\arraystretch{1.15}
\caption{Experimental results of identifying NAT devices on the Internet.}
\vspace{-2mm}
\begin{threeparttable}
\resizebox{0.95\linewidth}{!}{%
\begin{tabular}{c|ll|ll|ll|ll|ll|ll|ll|lll}
\bottomrule
 &
  \multicolumn{2}{c|}{\textbf{Frankfurt}} &
  \multicolumn{2}{c|}{\textbf{Virginia}} &
  \multicolumn{2}{c|}{\textbf{California}} &
  \multicolumn{2}{c|}{\textbf{Jakarta}} &
  \multicolumn{2}{c|}{\textbf{Bangkok}} &
  \multicolumn{2}{c|}{\textbf{Beijing}} &
  \multicolumn{2}{c|}{\textbf{São Paulo}} &
  \multicolumn{3}{c}{\textbf{Total}} \\ \hline
\textbf{Request} &
  \multicolumn{2}{c|}{14,487} &
  \multicolumn{2}{c|}{14,428} &
  \multicolumn{2}{c|}{15,690} &
  \multicolumn{2}{c|}{9,102} &
  \multicolumn{2}{c|}{13,114} &
  \multicolumn{2}{c|}{10,328} &
  \multicolumn{2}{c|}{14,312} &
  \multicolumn{3}{c}{91,461} \\ \hline
\textbf{Denial} &
  \multicolumn{2}{c|}{7,936} &
  \multicolumn{2}{c|}{8,235} &
  \multicolumn{2}{c|}{9,617} &
  \multicolumn{2}{c|}{4,808} &
  \multicolumn{2}{c|}{8,001} &
  \multicolumn{2}{c|}{5,832} &
  \multicolumn{2}{c|}{7,370} &
  \multicolumn{3}{c}{51,799} \\ \hline
\textbf{Approval} &
  \multicolumn{2}{c|}{6,551} &
  \multicolumn{2}{c|}{6,193} &
  \multicolumn{2}{c|}{6,073} &
  \multicolumn{2}{c|}{4,294} &
  \multicolumn{2}{c|}{5,113} &
  \multicolumn{2}{c|}{4,496} &
  \multicolumn{2}{c|}{6,942} &
  \multicolumn{3}{c}{39,662} \\ \hline
\textbf{Clients} &
  \multicolumn{2}{c|}{5,616} &
  \multicolumn{2}{c|}{5,045} &
  \multicolumn{2}{c|}{3,458} &
  \multicolumn{2}{c|}{3,536} &
  \multicolumn{2}{c|}{3,883} &
  \multicolumn{2}{c|}{3,184} &
  \multicolumn{2}{c|}{5,432} &
  \multicolumn{3}{c}{30,154} \\ \hline
\rowcolor[HTML]{EFEFEF} 
\cellcolor[HTML]{EFEFEF} &
  \multicolumn{1}{c|}{\cellcolor[HTML]{EFEFEF}} &
  486A &
  \multicolumn{1}{c|}{\cellcolor[HTML]{EFEFEF}} &
  386A &
  \multicolumn{1}{c|}{\cellcolor[HTML]{EFEFEF}} &
  275A &
  \multicolumn{1}{c|}{\cellcolor[HTML]{EFEFEF}} &
  316A &
  \multicolumn{1}{c|}{\cellcolor[HTML]{EFEFEF}} &
  334A &
  \multicolumn{1}{c|}{\cellcolor[HTML]{EFEFEF}} &
  275A &
  \multicolumn{1}{c|}{\cellcolor[HTML]{EFEFEF}} &
  491A &
  \multicolumn{1}{c|}{\cellcolor[HTML]{EFEFEF}} &
  \multicolumn{1}{c|}{\cellcolor[HTML]{EFEFEF} 1,289A} &
  \cellcolor[HTML]{EFEFEF} \\ \cline{3-3} \cline{5-5} \cline{7-7} \cline{9-9} \cline{11-11} \cline{13-13} \cline{15-15} \cline{17-17}
\rowcolor[HTML]{EFEFEF} 
\multirow{-2}{*}{\cellcolor[HTML]{EFEFEF}\textbf{NAT}} &
  \multicolumn{1}{c|}{\multirow{-2}{*}{\cellcolor[HTML]{EFEFEF} 1,416}} &
  84C &
  \multicolumn{1}{c|}{\multirow{-2}{*}{\cellcolor[HTML]{EFEFEF} 1,158}} &
  87C &
  \multicolumn{1}{c|}{\multirow{-2}{*}{\cellcolor[HTML]{EFEFEF} 804}} &
  68C &
  \multicolumn{1}{c|}{\multirow{-2}{*}{\cellcolor[HTML]{EFEFEF} 927}} &
  66C &
  \multicolumn{1}{c|}{\multirow{-2}{*}{\cellcolor[HTML]{EFEFEF} 994}} &
  75C &
  \multicolumn{1}{c|}{\multirow{-2}{*}{\cellcolor[HTML]{EFEFEF} 863}} &
  68C &
  \multicolumn{1}{c|}{\multirow{-2}{*}{\cellcolor[HTML]{EFEFEF} 1,443}} &
  84C &
  \multicolumn{1}{c|}{\multirow{-2}{*}{\cellcolor[HTML]{EFEFEF} 7,605}} &
  \multicolumn{1}{c|}{\cellcolor[HTML]{EFEFEF} 124C} &
  \multirow{-2}{*}{\cellcolor[HTML]{EFEFEF}25.22\%} \\ \hline
\textbf{Separate IP} &
  \multicolumn{2}{c|}{2,449} &
  \multicolumn{2}{c|}{2,408} &
  \multicolumn{2}{c|}{1,650} &
  \multicolumn{2}{c|}{1,636} &
  \multicolumn{2}{c|}{1,819} &
  \multicolumn{2}{c|}{1,361} &
  \multicolumn{2}{c|}{2,425} &
  \multicolumn{2}{c|}{13,748} &
  45.59\% \\ \hline
\textbf{Unknown} &
  \multicolumn{2}{c|}{1,751} &
  \multicolumn{2}{c|}{1,479} &
  \multicolumn{2}{c|}{1,004} &
  \multicolumn{2}{c|}{973} &
  \multicolumn{2}{c|}{1,070} &
  \multicolumn{2}{c|}{960} &
  \multicolumn{2}{c|}{1,564} &
  \multicolumn{2}{c|}{8,801} &
  29.19\% \\ 
\toprule
\end{tabular}
}
\begin{tablenotes}
       \footnotesize
       \item[] ``A'' means ASes where the identified NAT devices reside, and ``C'' means countries that the identified NAT devices belong to.
    \end{tablenotes}
\end{threeparttable}
\vspace{-3mm}
\label{tab:NAT_identifying}
\end{table*}

\subsection{NAT Mappings Manipulation}

%
Table~\ref{NAT_removing} presents the findings of our experiments on off-path removal of NAT mappings in various NAT setups. These setups encompass 6 native OSes, with 2 found to be vulnerable, 8 types of router firmware, 6 of which exhibit vulnerabilities, and 30 commercial routers from 14 vendors, 29 of which display vulnerabilities. The malicious removal against NAT mappings is achieved using crafted TCP \texttt{RST} packets with arbitrary sequence numbers.
Our experimental results demonstrate the disparities in NAT implementations among native OSes, various router firmware versions, and commercial routers. Particularly noteworthy is the vulnerability exhibited by the majority of downstream real-world NAT devices\footnote{After we reported the vulnerability, the FreeBSD and OpenWrt communities promptly acknowledged it. Specifically, TCP sequence number validation on inbound packets is now enabled by default in pf(4) starting from FreeBSD 14.0 and in netfilter starting from OpenWrt 23.05, respectively.}.

\subsection{Case Study} \label{subsec:case-costs}


\noindent\textbf{i) DoS over SSH Connections.}
As shown in Fig.~\ref{pic:setup},  4 NATed clients behind the NAT device have established SSH connections with the victim server (our VPS deployed in Tencent Cloud and ALICLOUD,  respectively). These clients periodically send messages to the server. Additionally, other NATed clients accessing the NAT network may intend to establish new SSH connections with the server.
The attacker on the Internet utilizes the method illustrated in Fig.~\ref{pic:dos} to launch a DoS attack, aiming to terminate SSH connections already established by the 4 NATed clients and disrupt the establishment of new SSH connections, thereby blocking the NAT network from connecting to the SSH server.
The attacker continuously alternates between sending multiple crafted TCP \texttt{RST} packets (up to 65,535) to the NAT device, aiming to clear the session mappings, and sending crafted \texttt{PUSH/ACK} packets to the server to disrupt the SSH connections. This orchestrated sequence of actions constitutes a DoS attack.

\begin{figure}[h]
    \vspace{-2mm}
    \centering
    \includegraphics[width=0.8\linewidth]{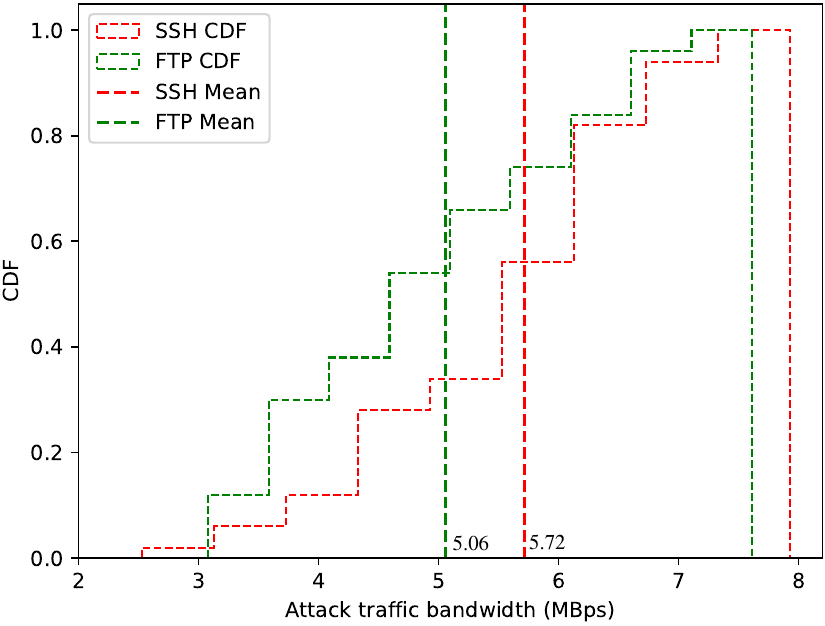}
    \vspace{-2mm}
    \caption{Attack traffic bandwidth of DoS over SSH and FTP.}
    \label{pic:cdf-ssh-ftp}
    \vspace{-1mm}
\end{figure}

Fig.~\ref{pic:cdf-ssh-ftp} presents the empirical cumulative distribution function (CDF) of the attack traffic bandwidth. We conduct 50 experiments, and our experimental results reveal that using an average bandwidth of 5.72MBps, the off-path attacker can block all NATed clients behind the vulnerable NAT device from connecting to the SSH server (including hindering the establishment of new SSH connections).
\major{
The crafted TCP \texttt{RST} packets and \texttt{PUSH/ACK} packets sent to the NAT device and the server, respectively, do not trigger alerts on either the NAT device or the server. Notably, out-of-order \texttt{PUSH/ACK} packet transmission is common in practice. We observe that the tested server (running Linux, as well as Windows, FreeBSD, and macOS), secured by Tencent Cloud or ALICLOUD firewalls, does not block the crafted \texttt{PUSH/ACK} packets.}

\noindent\textbf{ii) DoS over FTP Downloads.}
Similarly, using our attack method, the attacker can terminate established TCP connections between the NATed clients and the FTP server in clouds or prevent the establishment of new connections.
This obstruction hinders communication between the NAT network and the FTP server, leading to disruptions in FTP file downloads. We conduct 50 experiments and the average attack traffic bandwidth is about 5.06MBps, as shown in Fig.~\ref{pic:cdf-ssh-ftp}.

\section{Real-World Attacks}
\label{sec:real-world}



In addition to end-to-end evaluations, we also conduct a comprehensive study to evaluate the impacts of our attacks in the real world. The results of our ethical experiments on the Internet show that our attacks could significantly damage the real world.
Firstly, over an 11-month period, we identify more than 7,600 NAT devices that are distributed across 1,289 ASes in 124 countries around the world. 
Moreover, we conduct evaluations on 180 real-world NAT networks, including various scenarios such as public Wi-Fi networks, 4G LTE/5G networks, and cloud networks. The experimental results demonstrate that more than 92\% of the tested NAT networks are vulnerable to our DoS attack.
\major{
Note that even though our experiments involve human interactions, we fully addressed the ethical issues. We provided detailed explanations of the experiment’s purpose and methodology and obtained participants’ approval before conducting the experiments. Additionally, our experiments do not cause harm to the participants, and we provide them with reports on the experiment results (see Ethical Considerations in \S\ref{sec:intro} for more details).
}

\subsection{Identifying NAT Devices on the Internet}

\noindent\textbf{Experimental Setup.} 
We deploy 7 vantage points in 5 ASes, i.e., HTTP servers equipped with Apache 2.4.55/Linux 6.2.0, located globally in Frankfurt, Virginia, California, Jakarta, Bangkok, Beijing, and São Paulo.
We use the method in Fig.~\ref{pic:identifying_NATed_Networks} to determine whether the client requesting for our vantage point is a NATed client or a separate IP host.
We share the URLs for accessing our vantage points in social media platforms (e.g., TikTok and WeChat) and community forums (e.g., Dev community) for seeking voluntary users to participate in our NAT identification.
On our vantage points, we open TCP port 80 to listen for incoming requests. When an HTTP request arrives, we first obtain the user's approval before conducting the identification (see \S\ref{sec:intro} for our ethical considerations and Fig.~\ref{pic:NAT_snap} in Appendix for the snapshots of accessing our vantage points\footnote{In practice, the identification can be carried out without the awareness of the clients.}). Then, we change the path MTU between the user's machine and our vantage point. After that, we send probing packets and observe the size of the response packets to determine whether the user's machine is a separate IP host or a NATed client. If the user's machine is a NATed client, the source IP address of the request corresponds to a public IP address employed by the NAT device. 
%


\noindent\textbf{Experimental Results.} 
Table~\ref{tab:NAT_identifying} shows the results of our identification for NAT devices from the 7 vantage points.
At different vantage points, the experimental results may vary. For instance, at the vantage point in Frankfurt (the second column of Table~\ref{tab:NAT_identifying}), we receive a total of 14,487 HTTP requests over a 11-month period, i.e., from May 13, 2023, to April 13, 2024.
Out of the 14,487 requests, 7,936 deny our identification, while the remaining 6,551 approve. After eliminating duplicates within the 6,551 approved requests, 5,616 unique clients with different IP address access our HTTP server deployed at this vantage.
Among these 5,616 clients, our method reveals that 1,416 are located behind a NAT device, indicating the identification of 1,416 publicly routable IP addresses used by NAT devices. These NAT devices are spread across 486 ASes (i.e., 486A) in 84 countries (i.e., 84C).
2,449 out of the 5,616 clients are operating as separate hosts with unique IP address. The remaining 1,751 clients fall under the ``Unknown'' category due to undetectable results.
This implies that the reflected ICMP reply packets, responding to the vantage point's probing, are either blocked at the client or by middleboxes on the Internet~\cite{raman2020measuring}. As a result, it is difficult for the vantage point to determine whether these clients are within NAT networks or functioning as separate IP hosts.
%

We also encounter scenarios where clients might access our HTTP vantage points via a VPN proxy, making it challenging for the vantage points to distinguish whether the client is behind a NAT device or a separate IP host.
This challenge stems from the fact that when the vantage point sends an ICMP ``Fragmentation Needed and DF Set'' message to the client (in practice, this message is directed to the VPN proxy rather than the real source client) to adjust the client's path MTU (as shown in Fig.~\ref{pic:identifying_NATed_Networks}), our observations indicate that the VPN proxy typically discards the ICMP message instead of forwarding it to the actual source client, unlike a typical NAT device\footnote{In our end-to-end experimental setup, we further confirm that a VPN proxy does not forward ICMP ``Fragmentation Needed and DF Set'' messages, as exemplified by the well-known OpenVPN with version 2.5.9.
\major{Note that our setup confirms that VPN proxies obstruct the NAT identification. However, we cannot pinpoint VPN proxies specifically using this method, as other middleboxes might also drop ICMP messages.}
}.
Consequently, the ICMP ``Fragmentation Needed and DF Set'' message fails to reduce the client's path MTU value and thus affects the subsequent observations. Under this circumstance, the vantage point cannot distinguish whether the client is a separate IP host or a NATed client, and we classify this circumstance as ``Unknown''.

\begin{figure}[h]
	\vspace{-2mm}
	\begin{center}
		\includegraphics[width=0.49\textwidth]{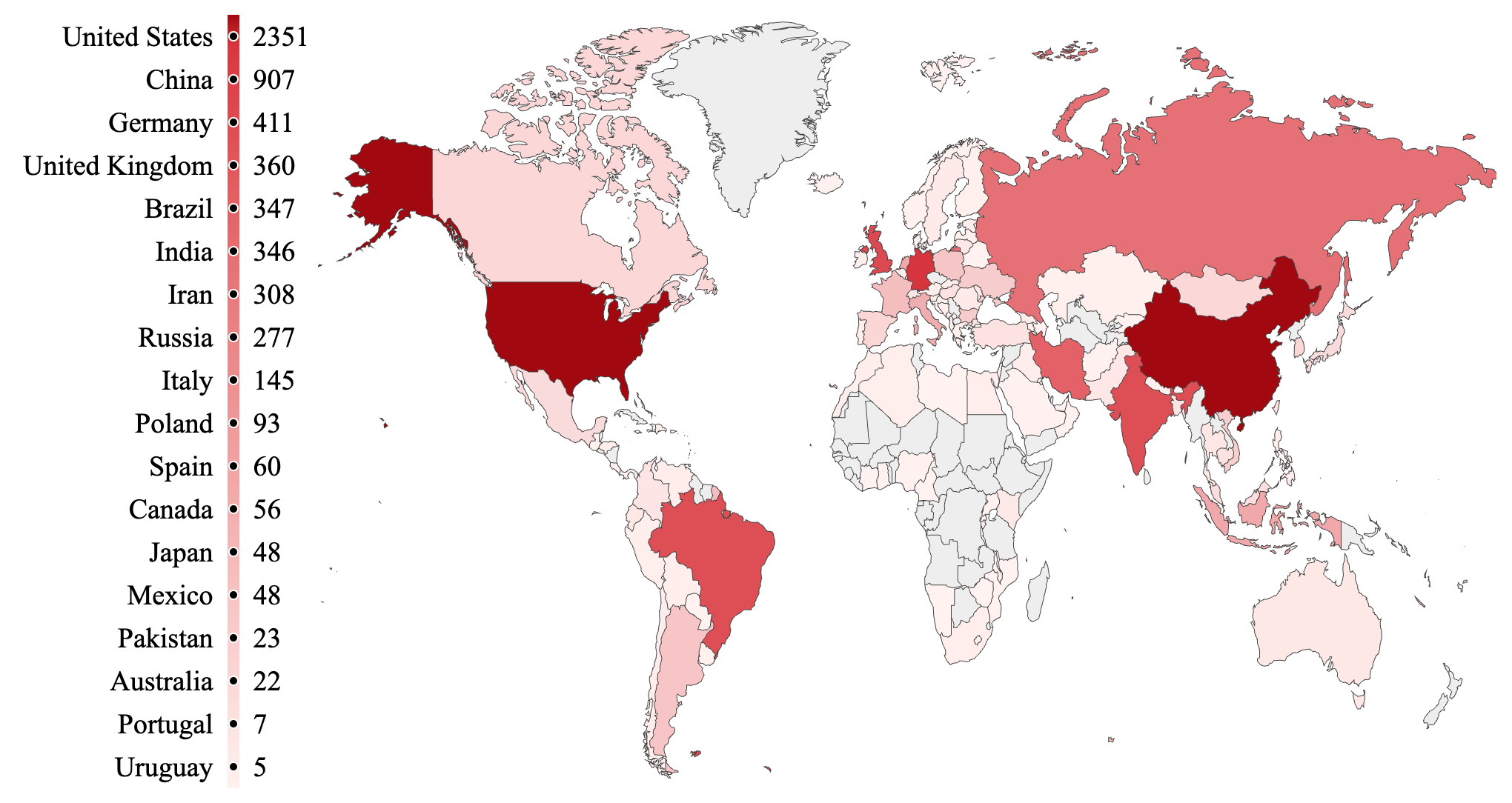}
		\vspace{-6mm}
		\caption{Distribution of the identified NAT devices.}
		\label{geo-NAT}
	\end{center}
	\vspace{-2mm}
\end{figure}

By aggregating the results from all 7 vantage points (as shown in the last column of Table~\ref{tab:NAT_identifying}), we determine that over a 11-month period, we receive a total of 91,461 HTTP requests\footnote{If the same source IP address requests our different vantage points, its HTTP requests will be recorded and handled only by the first vantage point requested, to avoid duplication.}. Out of the 30,154 clients, 7,605 (25.22\%) are identified as residing behind a NAT device. The identified NAT devices with public IP addresses are distributed across 1,289 ASes in 124 countries worldwide. 
Out of the remaining clients, 13,748 (45.59\%) are separate IP hosts, and the status of the remaining 8,801 (29.19\%) clients is unknown.
Table~\ref{NAT_detail} provides the details about 21 public IPv4 addresses used by identified NAT devices.
For instance, as illustrated in the first row, we identify a NAT device with a public IP address of ``*.145.177.*'' within a /24 CIDR range from the vantage point in Frankfurt. This NAT device is located in Jakarta, Indonesia.
Fig.~\ref{geo-NAT} shows the geographical distribution of all the identified NAT devices.

\begin{table}[h]
\caption{Details of 21 public IPv4 addresses used by NAT.}
\vspace{-3mm}
\begin{center}
\resizebox{0.8\linewidth}{!}{%
\begin{tabular}{@{}lclc@{}} 
			\toprule
\textbf{Public IP } &  \textbf{\begin{tabular}[c]{@{}c@{}} CIDR \end{tabular}} & \textbf{\begin{tabular}[c]{@{}c@{}} Location \end{tabular}}  & \textbf{Vantage Point} \\  \midrule
*.145.177.* & /24  &  Jakarta, Indonesia       & \multirow{3}{*}{Frankfurt} \\

*.74.192.* & /20 &  London, UK                &                            \\

*.52.158.* & /24  & Multan, Pakistan                &                            \\ \hline

*.147.62.* & /24  &  Kathmandu, Nepal                & \multirow{3}{*}{Virginia}     \\

*.190.220.* & /19 &  Donetsk, Russia               &                            \\

*.97.49.* & /20 &  Lima, Peru                &                            \\ \hline

*.64.76.* & /24  &  Kalemie, Congo                & \multirow{3}{*}{California}      \\

*.104.148.* & /19  &  Frankfurt, Germany               &                            \\

*.40.66.* & /21  & 	Xi'an, China               &                            \\ \hline

*.117.5.* & /19  &  Rasht, Iran                & \multirow{3}{*}{Jakarta}     \\

*.0.15.* & /16 &  Toledo, Spain                &                            \\

*.190.50.* & /20 &  Ramučiai, Lithuania     &                            \\ \hline

*.116.1.* & /18  &  Bistagno, Italy                & \multirow{3}{*}{Bangkok}      \\

*.64.125.* & /16  &  Banff, Canada               &                            \\

*.17.60.* & /24 &  Paris, France              &                            \\ \hline

*.113.106.* & /24 &  Palwal, India               & \multirow{3}{*}{Beijing}     \\

*.116.152.* & /12  &  Shenzhen, China               &                            \\

*.235.251.* & /24  &  Nova Prata, Brazil               &                            \\ \hline

*.171.124.* & /14  &  Lomas, Argentina                & \multirow{3}{*}{São Paulo}      \\

*.101.186.* & /24  &  Zalesye, Ukraine               &                            \\

*.199.82.* & /16 &   Grodzisk, Poland     &       \\ \bottomrule
		\end{tabular}
}
		\label{NAT_detail}
	\end{center}
\vspace{-4mm}
\end{table}

\subsection{Identifying Vulnerable NAT Networks}


%
%

%

\noindent\textbf{Experimental Setup.} 
Our attack involves four types of devices:
1) \textit{A victim client}, residing within the tested NAT networks, is under our control for ethical reasons. The victim client can be either a Xiaomi 12 cellphone or a VM rented from the tested clouds, equipped with a robust TCP/IP protocol suite implementation. Note that our attack does not impact regular users of the tested NAT networks.
2) \textit{A NAT device} responsible for translating and mapping internal clients' private IP addresses to public ones. This NAT device could be the gateway for public Wi-Fi networks, a PDN Gateway/UPF device in 4G LTE/5G networks, or the gateway for VMs in cloud networks.
3) \textit{An SSH server} located externally to the tested NAT networks. The server is deployed in California and is equipped with Linux 6.2.0 and OpenSSH 9.3.
4) \textit{An attack machine} equipped with Linux 5.19.0 and Scapy, with the capability of IP spoofing.
During our experiments, the victim client initially connects to the target NAT networks. It then establishes a TCP connection to the remote SSH server. The NAT device maintains a session mapping for the TCP connection between the victim client and the remote server. The off-path attacker will disrupt the SSH session between the victim client and the SSH server by exploiting the proposed DoS attack shown in Fig.~\ref{pic:dos}. 

\begin{figure}[h]
	\vspace{-2mm}
	\begin{center}
		\includegraphics[width=0.48\textwidth]{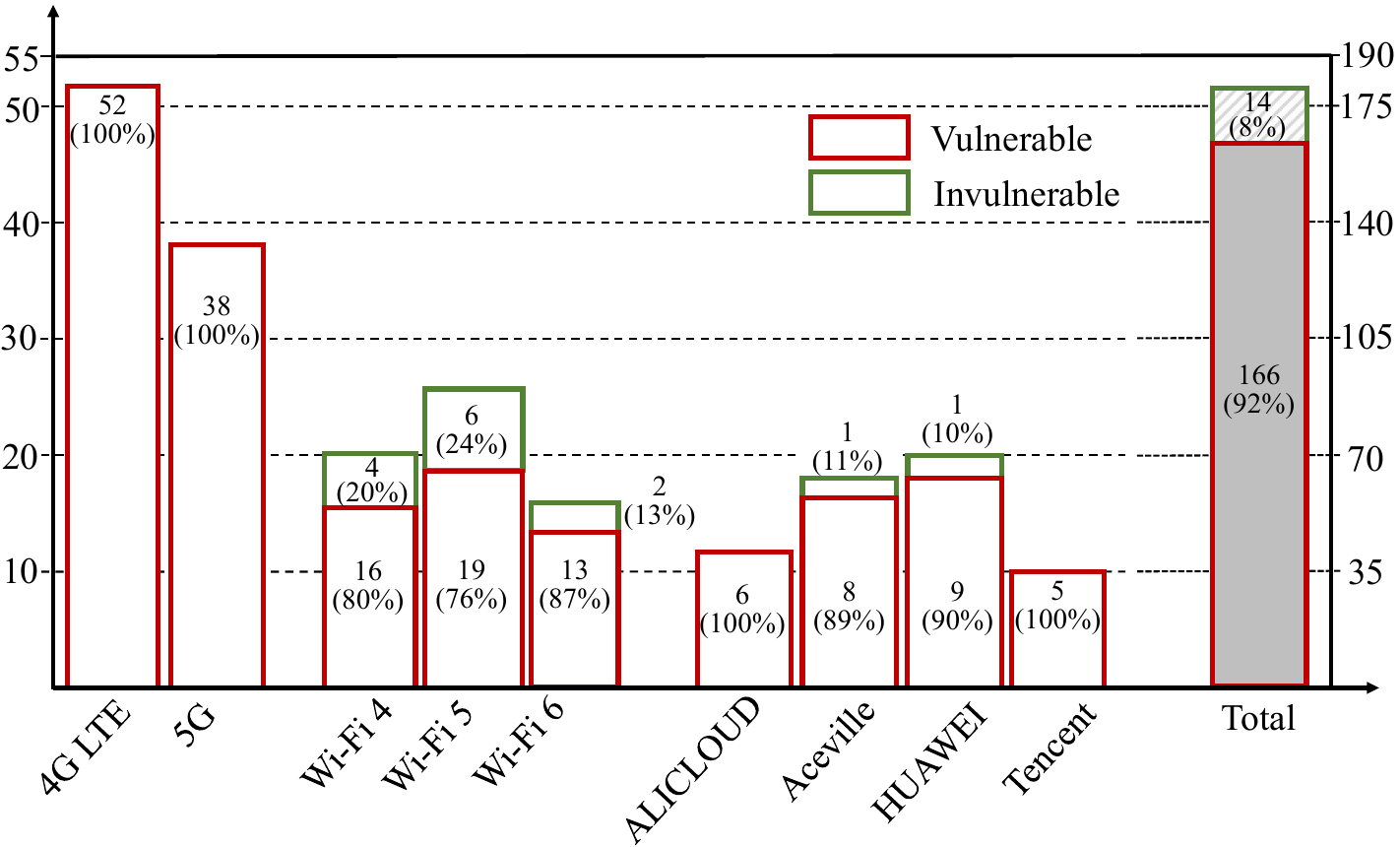}
		\vspace{-4mm}
		\caption{Evaluations on 180 real-world NAT networks.}
		\label{NAT-DoS-total}
	\end{center}
	\vspace{-2mm}
\end{figure}

\noindent\textbf{Experimental Results.}
\major{Given the widespread use of NAT in the real world, we randomly select 180 NAT networks from three popular network scenarios for a comprehensive evaluation. These networks include 90 4G LTE/5G networks, 60 Wi-Fi networks, and 30 cloud networks, all located in different regions.
}
The overall experimental results are illustrated in Fig.~\ref{NAT-DoS-total}.
Out of the 90 4G LTE/5G networks, all 52 4G LTE NAT networks and the remaining 38 5G NAT networks are vulnerable. The attacker can successfully terminate TCP connections initiated from cellphones connected to these networks.
In regard to the 60 Wi-Fi networks, our attack impacts 16 Wi-Fi 4 NAT networks, as well as 19 Wi-Fi 5 networks and 13 Wi-Fi 6 networks. As a result, the proportion of vulnerable NAT networks for Wi-Fi 4, Wi-Fi 5, and Wi-Fi 6 in our tests are 80\%, 76\%, and 87\%, respectively.

\begin{table*}[h]
\caption{Experimental results of our DoS attack over 30 real-world vulnerable NAT networks.}
\vspace{-4mm}
\begin{center}
\label{tab:NAT_DoS_details}
\begin{tabular}{@{}ccccllc@{}}
\toprule
\textbf{No.} & \multicolumn{1}{c}{\textbf{Public IP address}} & \textbf{CIDR} & \textbf{\begin{tabular}[c]{@{}c@{}}NAT\\ Scenario\end{tabular}} & \textbf{Region} & \multicolumn{1}{c}{\textbf{Organization}} & \textbf{\begin{tabular}[c]{@{}c@{}}Success\\ Rate\end{tabular}} \\ \midrule

1 & *.216.177.* & /20 & 4G LTE & Virginia, United States  & Verizon Business & 9/10 \\ 

2 & *.60.40.* & /16 & 4G LTE & Canarias, Spain  & VODAFONE ESPANA S.A.U. & 10/10 \\

3 & *.30.41.* & /24 & 4G LTE & Dhaka, Bangladesh  & Grameenphone Limited & 10/10 \\

4 & *.139.100.* & /19 & 4G LTE & Guizhou, China  & China Telecom  & 9/10 \\

5 & *.144.207.* & /21 & 4G LTE & Xinjiang, China  & China Mobile   & 10/10 \\

6 & *.254.3.* & /24 & 5G & Beijing, China  & China Unicom Beijing  & 10/10 \\

7 & *.108.164.* & /13 & 5G & Chongqing, China  & China Telecom  & 9/10 \\

8 & *.139.124.* & /15 & 5G & Shannxi, China   & China Unicom Shannxi  & 10/10 \\ 

9 & *.104.41.* & /22 & 5G & Guangdong, China & China Mobile  & 10/10 \\ 

10 & *.144.139.* & /23 & 5G & Sichuan, China & China Mobile  & 9/10 \\ \hline


\rowcolor[HTML]{EFEFEF}
11 & *.88.63.* & /18 & VM in cloud & California, United States  &  ALICLOUD & 10/10 \\
\rowcolor[HTML]{EFEFEF}
12 & *.74.95.* & /19 & VM in cloud & New South Wales, Australia  & ALICLOUD & 9/10 \\ 
\rowcolor[HTML]{EFEFEF}
13 & *.51.98.* & /22 & VM in cloud & Ontario, Canada  & Aceville & 10/10\\
\rowcolor[HTML]{EFEFEF}
14 & *.130.146.* & /19 & VM in cloud & Virginia, United States  & Aceville & 10/10\\ 
\rowcolor[HTML]{EFEFEF}
15 & *.135.216.* & /19 & VM in cloud & São Paulo, Brazil  & Aceville &  10/10\\ 
\rowcolor[HTML]{EFEFEF}
16 & *.163.199.* & /19 & VM in cloud & Tokyo, Japan  & Aceville & 10/10 \\ 
\rowcolor[HTML]{EFEFEF}
17 & *.138.165.* & /20 & VM in cloud & Johannesburg, South Africa  & HUAWEI CLOUDS & 9/10\\ 
\rowcolor[HTML]{EFEFEF}
18 & *.44.39.* & /20 & VM in cloud & Istanbul, Turkey  & HUAWEI CLOUDS & 10/10\\ 
\rowcolor[HTML]{EFEFEF}
19 & *.46.221.* & /17 & VM in cloud & Beijing, China  & HUAWEI CLOUDS &  10/10\\
\rowcolor[HTML]{EFEFEF}
20 & *.195.177.* & /18 & VM in cloud & Shandong, China  & Tencent Cloud & 10/10 \\ \hline


21 & *.36.245.* & /16 & Wi-Fi & Virginia, United States  & Verizon Business & 10/10 \\

22 & *.66.18.* & /16 & Wi-Fi & Virginia, United States  & Verizon Business & 10/10 \\

23 & *.198.141.* & /22 & Wi-Fi & Washington, United States  & Cox Communications Inc. & 10/10 \\

24 & *.223.36.* & /15 & Wi-Fi & California, United States & Comcast Cable Communications, LLC & 9/10 \\

25 & *.58.21.* & /16 & Wi-Fi & Burnaby, Canada  & Simon Fraser University & 9/10 \\

26 & *.138.139.* & /10 & Wi-Fi & Hesse, Germany  & Deutsche Telekom AG & 8/10 \\

27 & *.92.167.* & /20 & Wi-Fi & Kerala, India  & Bharat Sanchar Nigam LTD & 10/10 \\


28 & *.129.63.* & /18 & Wi-Fi & Beijing, China & China Unicom Beijing & 10/10 \\

29 & *.47.33.* & /24 & Wi-Fi & Dhaka, Bangladesh  & Link3 Technologies Limited & 10/10 \\

30 & *.114.95.* & /14 & Wi-Fi & Yunnan, China  & China Telecom  & 10/10 \\ \bottomrule



\end{tabular}
\vspace{-6mm}
\end{center}
\end{table*}

Moreover, we conduct tests on the NAT networks from 4 popular cloud providers, i.e., 6 ALICLOUD NAT networks, 9 Aceville NAT networks, 10 HUAWEI CLOUDS NAT networks, and 5 Tencent Cloud NAT networks. Out of these tested NAT networks, all 6 ALICLOUD NAT networks are vulnerable. Besides, our attack affects 8 Aceville NAT networks (i.e., 89\% of the tested networks). Similarly, 9 HUAWEI CLOUDS NAT networks experience the impact of our attack (i.e., with a proportion of 90\%). Lastly, all 5 Tencent Cloud NAT networks are affected by the attack.
In total, out of the 180 real-world NAT networks, 166 are vulnerable to our attack, indicating that the proportion of vulnerable NAT networks are more than 92\%. Table~\ref{tab:NAT_DoS_details} shows the details of 30 vulnerable real-world NAT networks in our evaluations. 
For example, as shown in the first row, we find that a 4G LTE NAT network (with a public IP address of ``*.216.177.*'' and CIDR of /20) located in Virginia, United States is affected by our attacks. This network belongs to Verizon Business.
In our testing, out of 10 attacks against this NAT network, 9 of them are successful. The one unsuccessful attempt is due to packet loss of the forged \texttt{RST} packet caused by network conditions.
\major{
It is worth noting that although our attack targets vulnerable NAT implementations and is independent of layer-2 protocols like Wi-Fi and LTE, the comprehensive evaluations show that a wide range of network access scenarios utilizing NAT are susceptible to our attacks.
}

\noindent \textbf{Reasons for Failures.}
The failure of our attacks on 14 tested NAT networks can be attributed to two situations.
%
%
Firstly, the crafted TCP \texttt{RST} packets fail to manipulate the preserved mappings in the NAT device of the target NAT networks. 
During our investigations on 60 public Wi-Fi networks, we identify 12 instances where failures occur due to this reason. In these instances, when the NAT device receives the crafted TCP \texttt{RST} packets, it refrains from responding by removing the corresponding mappings. Instead, it directly forwards the \texttt{RST} packets to the internal clients (which are eventually discarded by the clients due to the specified incorrect sequence number). We can recognize this situation by noting that the internal client receives the crafted \texttt{RST} packets issued by the attacker, but the attack fails. This indicates that the NAT device forwarded the \texttt{RST} packets directly and did not remove the preserved mappings.
Besides, our attack may also be foiled by specific network configurations, such as middle boxes or firewalls. For instance, during our investigations on cloud networks, we encounter 2 cloud NAT networks where our attack fails. In these cases, the crafted \texttt{RST} packets are blocked before they could remove the session mappings. We can identify this situation by observing that the internal client does not receive the crafted \texttt{RST} packets issued by the attacker, as well as the failure of our attack.
%
%

\section{Discussion and Countermeasure}
\label{sec:dis-counter}

\major{In this section, we discuss the impact of our attack on different NAT types and IPv6. We also propose countermeasures.}

\subsection{Impacts on Different NAT Types}

NAT technology, including various types like CGNAT (Carrier-Grade NAT), SNAT (Source NAT), DNAT (Destination  NAT),and NAT-enabled IoT CPE routers, is widely used in real world. 
Essentially, NAT relies on preserving session state to accurately route traffic between internal and external hosts. These NAT variants maintain a session mapping table to track active sessions, including TCP connections, and log communication between internal and external hosts. 
While the differences among various NAT types stem from their use cases, they all fundamentally center on managing the session mappings for routing network traffic.
Our attack leverages the TCP session mappings, enabling Internet attackers to craft TCP packets for manipulation. Consequently, the success of our attack hinges on the target NAT device maintaining stateful TCP connection mappings without enforcing legitimacy checks on the received TCP packets. This vulnerability may exist in NAT devices implementing various NAT types, including SNAT, DNAT, CGNAT, and others.

Through our empirical evaluations, we identify that SNAT and CGNAT devices, widely deployed in the real-world, exhibit this vulnerability significantly. A large number of SNAT routers (as shown in Table~\ref{NAT_removing}) and CGNAT devices from ISPs (as shown in Table~\ref{tab:NAT_DoS_details}) are affected by our attack\footnote{In our tests, certain CGNAT devices, e.g., China Unicom's CGNAT devices in Beijing (Table~\ref{tab:NAT_DoS_details} rows 6 and 28), maintain stateful TCP connection mappings without performing legitimacy checks, leaving the mappings vulnerable to malicious removals. Interestingly, these devices do not adhere to the TCP specifications for generating subsequent \texttt{RST} packets to tear down the victim server's sockets.
Nonetheless, even in this scenario, the DoS attack remains effective because the removal of the NAT device's mappings disrupts synchronization between the server and the NATed clients, resulting in dropped data and exceptions.}. 
Regarding DNAT, we establish an experimental environment using the OpenWrt 22.03 router firmware as a NAT device to enable the DNAT functionality. Our experimental results reveal that the DNAT implementation in OpenWrt 22.03 is vulnerable. Consequently, it is highly likely that many NAT devices with the DNAT feature in the real world are also susceptible. This is particularly concerning because OpenWrt serves as the foundational firmware for over 20 derivative projects\footnote{OpenWrt followed by more than 20 derivative projects provides firmware for over 2,000 types of NAT devices (\url{https://openwrt.org/toh/start}).}. Additionally, NAT-enabled IoT CPE routers may be at risk, as the vulnerability persists in many underlying router firmware. For instance, our testing shows that the ZTE 4G CPE 2 PRO router is vulnerable.

%

\subsection{\major{Impacts on IPv6}}
\major{
NAT is also widely used in IPv6 networks, including NAT64, NAT66, and NAT46. Therefore, we discuss the impact of our attack on these IPv6 NAT technologies. Using the widely adopted OpenWrt 22.03 firmware as a NAT device, we construct IPv6 networks for NAT64, NAT66, and NAT46, and test whether the two vulnerabilities revealed in this paper (i.e., the side channel in the PMTUD mechanism for NAT identification and the removal of NAT session mappings via crafted TCP \texttt{RST} packets) are still effective in these networks.
For the side channel vulnerability, we identify that this fundamental vulnerability consistently exists in three IPv6-related NAT scenarios. First, in our NAT64 network setup, an IPv6-enabled client accesses our server, deployed in Tencent Cloud, through the OpenWrt 22.03-equipped NAT64 device holding a public IPv4 address. The NAT64 device translates the address information from IPv6 space to IPv4 space and also translates the server's response back to IPv6 space. 
According to our method in Fig.~\ref{pic:identifying_NATed_Networks}, after the server issues the ICMPv4 ``Fragmentation Needed and DF Set'' message, the NAT64 device translates this message to an ICMPv6 ``Packet Too Big'' message (ICMPv6 error message with \texttt{Type}=2 and \texttt{Code}=0) and then forwards this ICMPv6 message to the client.
Consequently, the client updates its path MTU value to the server. From the server's perspective, the size of returned TCP packets is then reduced accordingly, while the ping reply packets from the same source remain unaffected. By contrast, path MTU desynchronization does not occur if the requester to the server is a separate IPv4 or IPv6 host.
Similarly, in our NAT46 and NAT66 network setups, the ICMPv6 ``Packet Too Big'' message issued by the server also triggers path MTU desynchronization, regardless of whether the NAT46 device forwards the translated ICMPv4 ``Fragmentation Needed and DF Set'' message to the IPv4 client or the NAT66 device directly forwards the ICMPv6 message to the IPv6 client.
}

\major{
With regard to the vulnerability of NAT session mapping removal, we also test three IPv6 NAT networks (i.e., networks linked by NAT64, NAT46, and NAT66 devices respectively, with the NAT devices equipped with the popular OpenWrt 22.03 firmware) to determine whether an off-path attacker impersonating the server and forging a TCP \texttt{RST} packet with an arbitrary sequence number can deceive the NAT device into removing the corresponding TCP session mappings with the server.
The experimental results show that all three NAT setups in our tests are vulnerable because the session mapping maintenance module within the NAT implementation does not enforce TCP sequence number checking. This vulnerability is independent of the specific IP protocol (i.e., regardless of whether it is IPv4 or IPv6).
}

\subsection{Countermeasures}\label{subsec:countermeasure}

\noindent \textbf{Responsible Disclosure.} 
We reported the side channel that can be exploited to identify NAT devices to IETF.
%
Currently, we are discussing this issue with the IETF security area directors, and we are told that the issue has been added as an item for an upcoming IETF Security Area meeting.
Besides, we reported our attack to the affected OS communities and identified ISPs. Following our disclosure, FreeBSD, China Telecom, China Unicom, China Mobile, ALICLOUD, HUAWEI CLOUDS, and Tencent Cloud confirmed the occurrence of our DoS attack stemming from their NAT implementations. These entities also recognized our efforts in enhancing the security of their services.
Furthermore, two prominent NAT firmware platforms, OpenWrt and Asuswrt, verified our DoS attack. The underlying vulnerability within the core NAT firmware extends its impact to a significant number of downstream NAT vendors, including NETGEAR, Linksys, Huawei, TP-Link, H3C, RuiJie, and Xiaomi, among others. 
We responsibly disclosed this vulnerability to the affected vendors and received acknowledgments from Linksys, Huawei, TP-Link, and Xiaomi. We obtained 5 CVE/CNVD identifiers (CVE-2023-6534, CVE-2023-31635, CNVD-2023-60783, CNVD-2023-30194, CNVD-2023-30193) for the disclosed vulnerabilities.
In addition, we recommend our countermeasures to prevent the identified DoS attack.

\noindent \textbf{Fixing the Side Channel in PMTUD.}
\major{The root cause of the side channel, which allows attackers on the Internet to remotely identify NAT devices, is a design flaw in NAT's PMTUD mechanism. Specifically, this flaw arises from insufficient consideration of the PMTUD mechanism in NAT specifications, leading to information leakage.}
This lack of synchronization in the path MTU values from the same source IP address results in exploitable information leakage, enabling the identification of NAT devices. NAT technology is extensively employed in diverse network scenarios, such as 4G LTE/5G, Wi-Fi, IoT, ICS, and others. Consequently, it is crucial to address this information leakage effectively and prevent any further attacks.
We propose enhancing the design of PMTUD for NAT networks by mandating that NAT devices not only translate ICMP ``Fragmentation Needed and DF Set'' messages to internal clients but also synchronize and update their own path MTU values for the server. This approach ensures the consistency of path MTU values from the same source IP address, effectively mitigating potential security risks arising from information leakage.
Note that NAT devices may face a challenge in verifying the legitimacy of the received ICMP error messages before updating their own path MTU values. Addressing this requires NAT devices to maintain more information, e.g., the window range of TCP connections, enabling the validation for the received ICMP error message's legitimacy. We have reported this issue to the IETF.

\noindent \textbf{Enforcing More Strict Checks on TCP.}
%
%
\major{
The vulnerability of NAT mapping removal arises from the absence of essential checks for TCP \texttt{RST} packets in NAT implementations across various scenarios.}
Although some recent OSes and router firmware have implemented security mechanisms to verify the legitimacy of received TCP \texttt{RST} packets, the majority of real-world NAT devices—such as commercial NAT routers—often do not perform these security checks. This oversight can potentially lead to DoS attacks.
A straightforward solution is to mandate the NAT devices implement more strict checks on the received TCP packets, particularly TCP control packets like \texttt{RST}, by verifying if the carried sequence number falls within the acceptable range of the corresponding TCP connection. 
%
%
Note that more strict validations may potentially introduce vulnerabilities, making the NAT device susceptible to attacks. Due to limited resources in NAT devices, attackers could exploit this by flooding the device with a high volume of forged \texttt{RST} packets, compelling it to conduct additional checks and depleting excessive resources. 
%
Nonetheless, we advocate for the implementation of more strict checks on the received TCP packets by NAT devices to mitigate potential attacks, especially considering the crucial role that NAT plays in the existing Internet infrastructure.
%
%
A prototype based on OpenWrt 22.03 confirms the effectiveness of our countermeasure by enabling TCP window checks, effectively thwarting remote attackers' manipulation of HTTP sessions issued from NATed clients.
\major{
We modify the kernel of our NAT device (running OpenWrt 22.03) to check the sequence number of received TCP packets before responding to them. This prevents off-path attackers from removing TCP session mappings by sending crafted out-of-band TCP \texttt{RST} packets with incorrect sequence numbers, thereby thwarting the DoS attack. As a result, HTTP access from our NATed clients to the remote HTTP server remains unaffected in our tests.
}

\section{Related Work}
\label{sec:related}
\major{In this section, we review previous related works from two aspects: NAT issues and DoS attacks.}

\noindent \textbf{NAT Issues.}
NAT plays a crucial role in network connectivity, making it a prominent focus for academic studies.
Prior works show that malicious insiders of NAT networks may consume limited resources of the NAT device to construct a DoS attack~\cite{rfc5382,nguyen2018slow,winemiller2012nat}. 
For example, a malicious NAT client may establish multiple (as many as 65,535) TCP connections with a target server,
thus preventing other clients from being assigned available ephemeral ports to initiate TCP connections to the server.
Distinctively, in our attack, the attacker is not limited by network topology, residing remotely on the Internet rather than within the local NAT network.
By rewriting NAT session mappings and then intercepting the victim client's TCP packets in Wi-Fi Networks, Yang \etal exploited sequence number leakage to hijack TCP connections~\cite{yangexploiting}. They also discussed constructing remote TCP DoS attacks.
%
However, a significant challenge for their DoS attack is the precise identification of a remote NAT-enabled Wi-Fi router. We uncover a fundamental side channel in the PMTUD mechanism of NAT specifications that can be exploited to remotely identify NAT devices on the Internet. Furthermore, we explore NAT implementation disparities across native OSes, various router firmware, and commercial routers, revealing our DoS attack's fundamental impact on a wide range of real-world NAT networks, i.e., beyond Wi-Fi to 4G LTE/5G cellular networks, Cloud VPS networks, and more.

Bellovin proposed a method to identify a NAT network and the associated hosts by examining the distributions of IP IDs originating from a particular IP address~\cite{bellovin2002technique}. However, this approach may encounter difficulties when dealing with packets that have zero or random IP IDs, leading to potential complications. 
Other OS features such as the Don't Fragment (DF) flag, the Time-To-Live (TTL) field, the TCP window size, the TCP source port, the initial sequence number (ISN), and the SYN packet size were also utilized to identify NAT networks and determine the number of hosts behind the network~\cite{beverly2004robust,sflow,OSprinting,NATleak,rfc6528,rfc3022}.
Essentially, the concept behind these methods is that if an IP address is associated with multiple hosts, it could suggest the presence of a NAT device with distinct OS features behind it. However, it is easy to see that these methods may generate substantial false alarms in situations where a computer has multiple OSes installed or when traffic obfuscation techniques are employed.

By exploiting middle-box protocols (e.g., UPnP), an attacker could connect devices behind NAT gateways to identify NAT networks. However, this technique requires the deployment of application-layer middle-box protocols or the enabling of malicious scripts on devices behind NAT gateways~\cite{rytilahti2020using}.
Gokcen \etal employed the Naive Bayes learning technique as a classifier to identify NAT devices based on given traffic traces. However, this approach heavily depends on the quality of the collected datasets~\cite{gokcen2014can}.
NAT was also exploited to execute DNS hijacking. For example, Herzberg \etal demonstrated techniques for bypassing source port randomization in NAT networks and hijacked a local DNS resolver~\cite{herzberg2012security}.
\noindent \textbf{DoS Attacks.}
DoS attacks and network traffic manipulations have been extensively studied in recent years~\cite{cho2019bgp,sermpezis2018survey,nakibly2014ospf,song2017novel}. Feng \etal discovered a side channel in the new mixed IP ID assignment policy that can be exploited by off-path attackers to terminate an SSH session~\cite{ccsfeng,feng2021off}.
Cao \etal, leveraging a side channel in the challenge ACK mechanism, uncovered a TCP hijacking attack capable of poisoning or resetting victim TCP connections~\cite{cao2018off,cao2016off}. Fortunately, most of these previous attacks have been addressed by the security community~\cite{ccsfeng,feng2021off,cao2016off,cao2018off}. 
Additionally, recent research has explored low-rate TCP-targeted DoS attacks~\cite{kuzmanovic2003low,jero2018automated,shevtekar2005low, herzberg2010stealth} and congestion-based attacks on intermediate links~\cite{smith2018routing,MuoiTran2019routing}.
In contrast to these studies, our research focuses on a covert DoS attack specifically targeting NAT networks. 
Previous works~\cite{TCP_Kill,killcx} demonstrated that TCP connections can be terminated by crafting \texttt{RST} packets, but they require the attacker to be physically present on the host to eavesdrop on the correct sequence number for crafting an acceptable \texttt{RST} packet. In contrast, our attack can be executed remotely by off-path attackers on the Internet.

\section{Conclusion}
\label{sec:conclusion}

%
In this paper, we conduct an empirical study on remote DoS attacks against NAT networks.
%
We uncover a side channel stemming from insufficient considerations of the PMTUD mechanism in the NAT specifications. This side channel leads to the remote identification of NAT devices on the Internet. 
%
Furthermore, we demonstrate that Internet attackers may manipulate TCP connection mappings in various NAT devices using carefully crafted TCP \texttt{RST} packets. This manipulation opens the door to remote DoS attacks targeting NAT networks. 
Through extensively empirical studies conducted on various NAT implementations and real-world NAT networks, we highlight the significance of our DoS attack.
%
Finally, we develop countermeasures against the attack.


\section*{Acknowledgment}
We thank the anonymous reviewers for their insightful comments.
%
This work was supported in part by the National Key Research and Development Program of China under Grant 2022YFB3102303, the National Science Foundation for Distinguished Young Scholars of China under No. 62425201, the Science Fund for Creative Research Groups of the National Natural Science Foundation of China under No. 62221003, the Key Program of the National Natural Science Foundation of China under No. 61932016 and No. 62132011.
Ke Xu is the corresponding author of this paper.



%

\bibliographystyle{IEEEtranS}
\bibliography{reference}

\appendix

\subsection{Snapshots of NAT Identification on the Internet}

\begin{figure}[h]
	\begin{center}
		\subfigure[Requiring the requester's approval to proceed with our experiment.]{ 
			\label{pic:snap_no}  
			\includegraphics[width=0.45\textwidth]{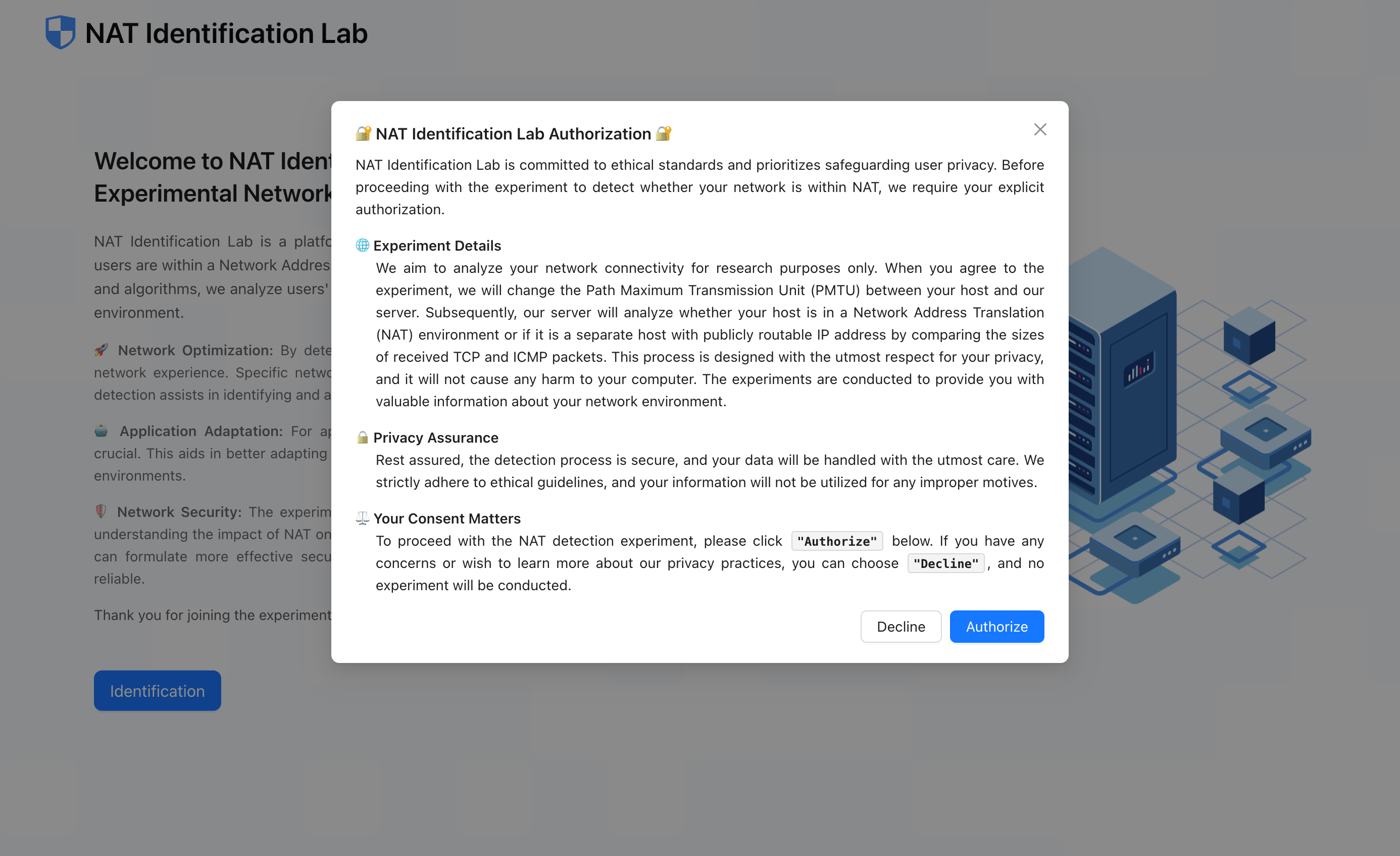} 
		} 
        \subfigure[The identification result is sent back to the separate host.]{ 
			\label{pic:snap_no}  
			\includegraphics[width=0.45\textwidth]{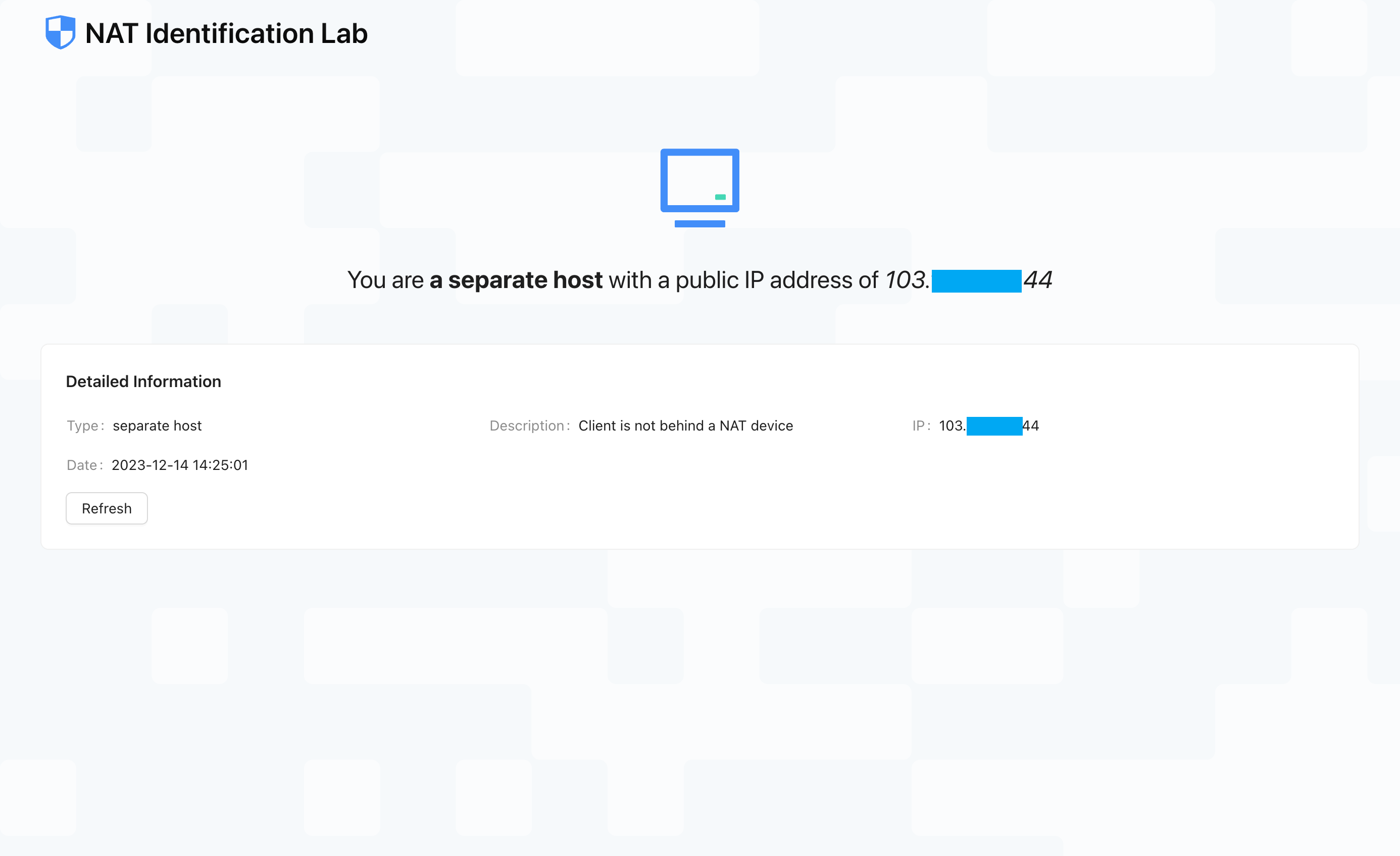} 
		} 
	\subfigure[The identification result is sent back to the NATed client.]{ 
			\label{pic:snap_yes}
			\includegraphics[width=0.45\textwidth]{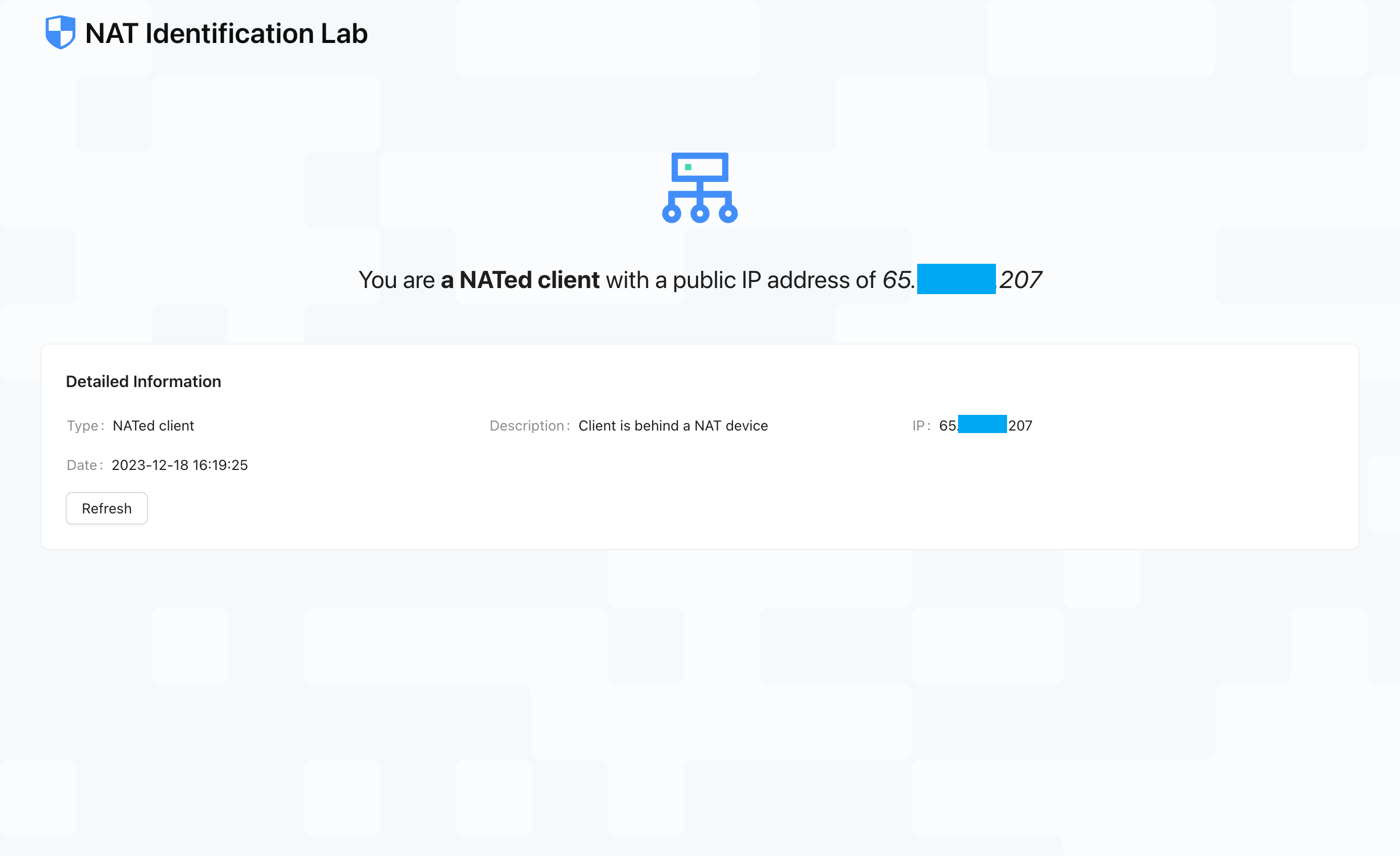}
		}
         \subfigure[The identification result is unknown.]{ 
			\label{pic:snap_yes}
			\includegraphics[width=0.45\textwidth]{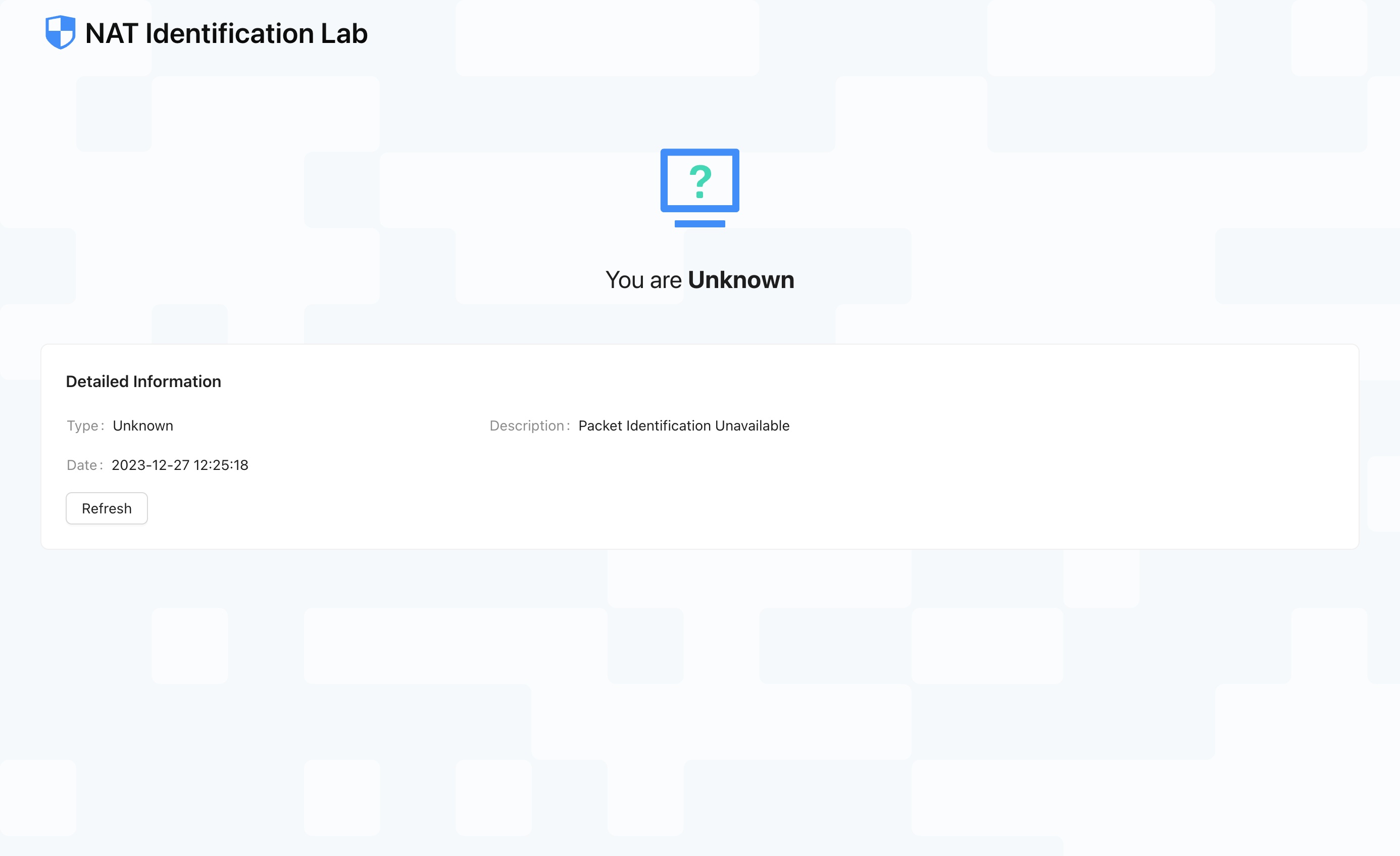}
		}
		\vspace{-2mm}
		\caption{Snapshots of NAT identification results, observed by the NATed client and the separate host in their web browsers, respectively, after clicking the URL to proceed with our experiment.} 
		\label{pic:NAT_snap} 
	\end{center}
\end{figure}

\end{document}